\documentclass[amsmath,amssymb,aps,twocolumn,superscriptaddress]{revtex4-2}
\usepackage{amsmath}
\usepackage{amssymb}
\usepackage{bm}
\usepackage{graphicx}
\usepackage{epsf}
\usepackage{xcolor}
\usepackage{color}
\usepackage{graphicx}% Include figure files
\usepackage[version=4]{mhchem}% change $\mathrm{SO_2}$ or SiO${_2}$ to -> \ce{SiO2}
\usepackage[bookmarks=false]{hyperref}
\usepackage{siunitx}% try to write down numbers and units in a consistent way
\usepackage[english]{babel}
\usepackage[T1]{fontenc}
\usepackage{dcolumn}% Align table columns on decimal point
\usepackage{tabularx}
\usepackage{array}

\newcommand{\q}{\mathbf q}
\newcommand{\rr}{\mathbf r}
\newcommand{\kk}{\mathbf k}
\newcommand{\vd}{\mathbf v_d}
\newcommand{\Tr}{\mathrm{Tr}}

\begin{document}

\title{Microscopic Theory of Drag in a Bose Condensate Interacting with a Moving Reservoir}
\author{Hassan Alnatah}
\thanks{Address correspondence to: alnatah@umd.edu}
\affiliation{Joint Quantum Institute, University of Maryland, 4254 Stadium Dr., College Park, Maryland 20742, USA}

\author{Shuang Liang}
\affiliation{Department of Physics, University of Pittsburgh, 3941 O’Hara Street, Pittsburgh, Pennsylvania 15218, USA}

\author{Shouvik Mukherjee}

\affiliation{Joint Quantum Institute, University of Maryland, 4254 Stadium Dr., College Park, Maryland 20742, USA}

\author{Qi Yao}
\affiliation{Joint Quantum Institute, University of Maryland, 4254 Stadium Dr., College Park, Maryland 20742, USA}
\author{Ashton S. Bradley}
\affiliation{Dodd-Walls Centre for Photonic and Quantum Technologies, Department of Physics, University of Otago, Dunedin, New Zealand}

\author{David W. Snoke}
\affiliation{Department of Physics, University of Pittsburgh, 3941 O’Hara Street, Pittsburgh, Pennsylvania 15218, USA}

\date{\today}
\begin{abstract}
We derive a microscopic theory of drag for a Bose-Einstein condensate interacting with a moving reservoir. Starting from interactions between a condensate and a drifting fermionic bath, we integrate out the bath degrees of freedom within the Born--Markov approximation to obtain an effective Gross--Pitaevskii equation with a drag potential. From this potential we derive an effective drag force and obtain a closed expression for the drag coefficient, determined by the reservoir density fluctuations and the condensate density profile. As an application, we simulate the drag for an exciton--polariton condensate interacting with a drifting electron gas.
\end{abstract}

\maketitle

\section*{Introduction}

Understanding how coherent quantum systems of bosons exchange momentum and energy with their environment is one of the central problems in many-body physics. In many physical systems, a Bose-Einstein condensate coexists with a noncondensed reservoir that can carry a finite flow, producing a moving environment that can exchange momentum with the condensate \cite{Meppelink2009}. Such situations arise in systems ranging from ultracold atomic gases interacting with thermal clouds \cite{Meppelink2009} to excitonic and polaritonic condensates interacting with free carriers \cite{Nandi2012,Myers2025}. The resulting drag between the condensate and its environment provides a mechanism for externally controlling the movement of the condensate \cite{Chervy2020,Myers2025}.

This notion of condensate drag deserves clarification. It is often stated in simplified terms that a superfluid does not experience drag. More precisely, a superfluid has no dissipative response to tangential forces, such as those exerted by the walls of a rotating annular trap on the condensate it confines \cite{Annett2004,NozieresPines1990,LifshitzPitaevskii1980}. A superfluid, however, will still react to body forces arising from momentum exchange with a dynamical reservoir, of the type considered here. 

Although drag phenomena have been investigated theoretically in many contexts -- exciton condensates and electrons \cite{Liu2017,Nguyen2025}, the drag of impurities moving through atomic Bose--Einstein condensates \cite{Astrakharchik2004,Sykes2009,frisch1992transition,pavloff2002breakdown}, and the drag of microcavity polaritons by free electrons \cite{Myers2025,Berman2010,Cotlet2019} --- their theoretical descriptions are often either kinetic or phenomenological. Kinetic treatments based on the semiclassical quantum Boltzmann equation \cite{Narozhny2016,Myers2025,Berman2010} evolve occupation numbers alone, and therefore carry no phase coherence --- precisely the property that defines a condensate. Phenomenological treatments operate at the opposite limit: effective drag forces or current-dependent potentials are introduced directly into nonlinear wave equations on the basis of symmetry arguments or fitted to experimental observations \cite{Ronning2020}. While both approaches successfully capture many qualitative features, neither reveals how the drag emerges from the underlying microscopic Hamiltonian, how its strength depends on the microscopic properties of the reservoir, or under what conditions local drag models are justified. A notable exception is the microscopic theory of Ref.~\cite{Cotlet2019}, which treats a degenerate Fermi sea at zero temperature and obtains the drag as a spatially uniform effective gauge (vector) potential --- a shift of the quasiparticle dispersion, equivalently a transconductivity; here, by contrast, the drag is derived for a general bath, with a Maxwell--Boltzmann distribution inserted only for evaluation, and appears as a position-dependent scalar potential in the real-space equation of motion of the coherent condensate.

A microscopic theory should begin from the many-body Hamiltonian and eliminate the reservoir degrees of freedom, yielding an effective theory for the condensate alone. Such a theory should determine the effective interaction with the reservoir without phenomenological parameters, and establish the conditions under which drag of the condensate emerges. More generally, it would provide a framework for describing the equations of motion for coherent quantum fluids interacting with their environments. A microscopic, open-systems framework of this kind already exists for atomic Bose gases in the form of the stochastic projected Gross--Pitaevskii equation, in which collisions with a thermal reservoir generate an energy-damping (scattering) term that acts as a real, number-conserving effective potential built from a scattering kernel~\cite{gardiner2003stochastic}. Unlike number damping, which transfers particles between the coherent field and reservoir, energy damping exchanges energy and momentum through number-conserving scattering. Its deterministic action is a real, generally nonlocal potential, allowing collective motion to relax without directly depleting the condensate; it therefore provides a highly coherent energy-transfer mechanism \cite{Rooney2012SPGPE}. This mechanism has since been investigated theoretically for finite-temperature nonlinear excitations. It provides a microscopic origin for vortex mutual friction and diffusion~\cite{Mehdi2023MutualFriction}, and governs the thermal decay of Jones--Roberts vortex dipoles and rarefaction pulses~\cite{Krause2024JRSolitons}. Despite these predictions and indirect quantitative comparisons, direct experimental tests of the energy-damping channel remain limited, and, more importantly, it has not yet been formulated for a condensate dragged by a distinct species, as considered here. 

In this work we develop such a framework. Starting from a microscopic interaction between a condensate and a moving fermionic bath, we integrate out the fermionic bath degrees of freedom and derive an effective Gross--Pitaevskii (G-P) equation for the condensate. We show that the entire influence of the environment is completely determined by an effective history-dependent potential acting on the condensate.

As a concrete application of the formalism, we consider an exciton--polariton condensate interacting with a moving electron gas. Exciton--polaritons have been shown to undergo Bose--Einstein condensation and exhibit long-range coherence in a wide range of experiments over the past two decades (e.g., \cite{deng2002condensation,kasprzak2006bose,balili2007bose,abbarchi2013macroscopic,sanvitto2010persistent,lagoudakis2009observation,amo2009superfluidity,yao2025persistent}). Exciton--polaritons provide an ideal platform because interactions between polaritons and electrons have been extensively studied experimentally \cite{ravets2019nonlinear,tan2020interacting,sidler2017fermi,ramon2002scattering}, including the demonstration of electrically controlled condensate motion \cite{Chervy2020,Myers2025}. Although we focus on this specific system, the derivation is general and may be applied to any coherent quantum fluid interacting with a moving reservoir.

\section*{Theory}

\begin{figure}
\includegraphics[width=1\columnwidth]{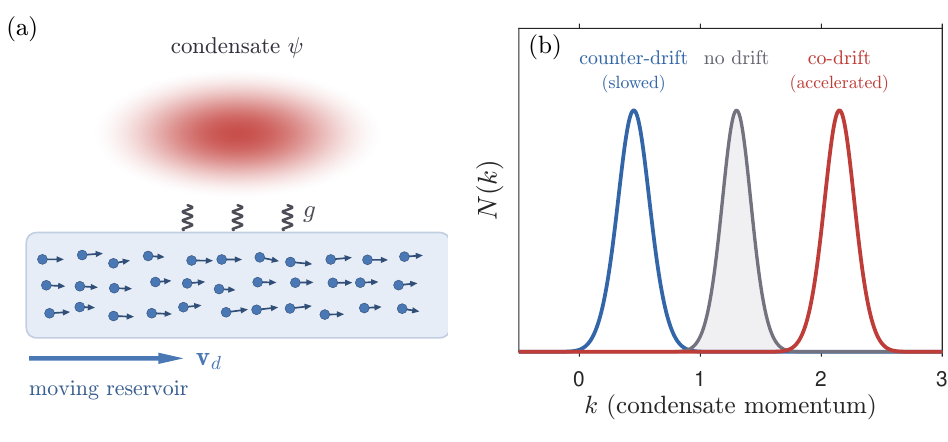}
\centering
\caption{Schematic of a condensate interacting with a moving reservoir. (a) A condensate $\psi$ interacting with a reservoir drifting with a velocity $\mathbf v_d$, producing a drag potential
$V_{\rm drag}$ acting on the condensate. (b) Momentum distribution $N(k)$ of the moving condensate: the reservoir
current accelerates it (co-drift) or slows it toward $k=0$ (counter-drift),
relative to no drift.}
\label{fig:one}
%%%%%%%%%%%%%%%%%%%%%%%%%%%
\end{figure}

We consider a Bose condensate coexisting with a reservoir that carries a uniform current, as shown in Fig.~\ref{fig:one}(a). The condensate and the reservoir interact with an interaction strength $g$; the condensate not only responds to the mean reservoir density, but also to the current of the reservoir. The aim is to eliminate the reservoir degrees of freedom and derive an effective equation of motion for the condensate alone, in which the influence of the moving reservoir manifests itself as an effective potential $V_{\rm drag}$. This potential can transfer momentum to the condensate, accelerating a co-propagating flow and slowing a counter-propagating one (Fig.~\ref{fig:one}(b)).

The treatment presented here is generic. We illustrate it with a nonequilibrium, ballistically moving condensate, as in Fig.~\ref{fig:one}(b), but the derivation nowhere assumes that the condensate is moving: the same effective equation governs an equilibrium, stationary condensate and a superfluid. The condensate must only be a coherent macroscopic field interacting with a reservoir.

We begin with the Hamiltonian $H=H_\psi+H_\phi+H_{\rm int}$ in $D$ spatial dimensions, where $H_\psi$ describes the bosons and $H_\phi$ describes the fermionic reservoir,
\begin{align*}
H_\psi
={}&
\int d^{D}\rr\,
\hat\psi^\dagger(\mathbf r)
\left[
-\frac{\hbar^2\nabla^2}{2m_\psi}
+
V_{\rm ext}(\mathbf r)
\right]
\hat\psi(\mathbf r)
\\
&+
\frac{U}{2}
\int d^{D}\rr\,
\hat\psi^\dagger(\mathbf r)
\hat\psi^\dagger(\mathbf r)
\hat\psi(\mathbf r)
\hat\psi(\mathbf r),
\\[1ex]
H_\phi
={}&
\int d^{D}\rr\,
\hat\phi^\dagger(\mathbf r)
\left[
\frac{-\hbar^2\nabla^2}{2m_\phi}
\right]
\hat\phi(\mathbf r).
\end{align*}

and $H_{\rm int}$ describes the boson-reservoir interaction
\begin{align*}
H_{\rm int}
&=
g
\int d^{D}\rr\,
\hat n_\psi(\mathbf r)\,
\hat n_\phi(\mathbf r),
\\
\hat n_\psi(\mathbf r)
&=
\hat\psi^\dagger(\mathbf r)\hat\psi(\mathbf r),
\qquad
\hat n_\phi(\mathbf r)
=
\hat\phi^\dagger(\mathbf r)\hat\phi(\mathbf r),
\end{align*}
where $U$ and $g$ are the strengths of the boson--boson and boson--reservoir contact interactions, respectively, and $\hat{\psi}$ and $\hat{\phi}$ are the bosonic and fermionic field operators, respectively. We note that in the stochastic projected Gross--Pitaevskii treatment an energy cutoff must be introduced to separate the coherent region from the thermal reservoir \cite{gardiner2003stochastic,blakie2008dynamics}, a choice that becomes essential once energy-damping interactions are included. No such mode cutoff is required here: the reservoir is a physically distinct species (the fermionic gas), so the system--reservoir separation is by particle species rather than by energy. The condensate is treated as a damped coherent field. We focus on the leading order effects of the reservoir, making a low-temperature approximation where the associated noise is neglected relative to the characteristic energy scales of the coherent field.

The system can be described by a density matrix $\rho$, which evolves according to the Liouville--von Neumann equation

\[
\frac{\partial \rho}{\partial t}
=
-\frac{i}{\hbar}[H,\rho],
\]
where $\rho$ contains the full quantum state of the combined condensate--reservoir system.

It is convenient to work in the interaction picture with respect to $H_0=H_\psi+H_\phi$. Transforming $\tilde H(t)=e^{iH_0t/\hbar}He^{-iH_0t/\hbar}$ and the density matrix $\tilde \rho (t)=e^{iH_0t/\hbar}\rho\, e^{-iH_0t/\hbar}$ gives
\begin{equation}
\frac{\partial \tilde \rho}{\partial t}
=
-\frac{i}{\hbar}
[
\tilde H_{\rm int}(t),
\tilde\rho(t)
].
\label{eq:von}
\end{equation}
Integrating \eqref{eq:von} with respect to $t$ and substituting the result back into itself gives
\begin{align}
\frac{\partial \tilde \rho}{\partial t}
={}&
-\frac{i}{\hbar}
[
\tilde H_{\rm int}(t),
\tilde\rho(0)
]
\notag\\
&-
\frac{1}{\hbar^2}
\int_0^t
dt'
[
\tilde H_{\rm int}(t),
[
\tilde H_{\rm int}(t'),
\tilde\rho(t')
]
].
\label{eq:iter}
\end{align}
\par
It is convenient to decompose the reservoir density as $\hat n_\phi = n_0 + \delta\hat n_\phi$, where $n_0 = \langle \hat n_\phi \rangle$ is the average density and $\delta\hat n_\phi$ is the fluctuations of the reservoir density obeying $\langle \delta \hat n_\phi \rangle$ = 0. This then allows us to write
\begin{align*}
\tilde H_{\rm int}(t)
&=
g
\int d^{D}\rr\,
\tilde n_\psi(\rr,t)\,
\left(n_0+\delta\tilde n_\phi(\rr,t)\right)
\\
&= g\,n_0\hat N_\psi+g\int d^{D}\rr\,
\tilde n_\psi(\rr,t)\,
\delta\tilde n_\phi(\rr,t),
\end{align*}
The first term is simply a constant energy shift of the bosons, which can be absorbed into the definition of $H_\psi$.

To evaluate the commutators in Eq.~\eqref{eq:iter}, we make three  approximations. First, we invoke the Born approximation \cite{breuer2002theory,carmichael2013statistical}, which assumes the boson--reservoir interaction is sufficiently weak so that the reservoir remains very close to its stationary drifting state throughout the evolution, allowing us to write $\tilde\rho(t)\simeq\tilde\rho_\psi(t)\otimes\rho_\phi$. Second, we invoke the Markov approximation for the reservoir, which assumes that the future evolution of a reservoir depends only on its current state, not on its past history. In other words, we assume short reservoir memory such that the reservoir correlations decay on a time scale $\tau_c$ during which the boson state barely changes, allowing us to replace $\tilde\rho_\psi(t')\to\tilde\rho_\psi(t)$ inside the $dt'$ integral. Importantly, this Markov approximation only assumes short memory of the reservoir{-}{-}it does not assume that the condensate itself is memoryless. Third, we assume that the reservoir is spatially uniform and in steady state.

\par
Using the approximations listed above, the first term in Eq.~(\ref{eq:iter}) vanishes when tracing over the reservoir since it is proportional to $\langle \delta \hat n_\phi \rangle$, which by definition averages to zero. The second term is nonzero and is proportional to $g^2$. After tracing over the reservoir, we obtain (details of the calculations are given in Appendix \ref{app:master})
\begin{align}
\frac{\partial \tilde \rho_\psi(t)}{\partial t}
={}&
-
\frac{g^2}{\hbar^2}
\int_0^\infty
d\tau
\int
d^{D}\rr\,d^{D}\rr'
\,
\Big\{
C(\rr-\rr',\tau)
\notag\\
&\times
\left[
\tilde n_\psi(\rr,t),
\tilde n_\psi(\rr',t-\tau)
\tilde\rho_\psi(t)
\right]
+\mathrm{h.c.}
\Big\},
\label{eq:ME}
\end{align}
where $C(\rr-\rr',t-t') = \langle \delta\tilde n_\phi(\rr,t) \, \delta\tilde n_\phi(\rr',t') \rangle $ and $\tau = t-t'$. After tracing over the reservoir, the only thing the bosons can know about the reservoir is the single function $C(\rr,\tau)$, which determines the drag due to the reservoir back-action on the condensate.

Transforming \eqref{eq:ME} back to the Schr\"odinger picture (the steps are given in Appendix~\ref{app:schroedinger}), gives
\begin{align}
\frac{\partial \rho_\psi(t)}{\partial t}
={}&
-\frac{i}{\hbar}\left[H_\psi,\rho_\psi\right]
\notag\\
&-
\frac{g^2}{\hbar^2}
\int_0^\infty
d\tau
\int
d^{D}\rr\,d^{D}\rr'
\,
\Big\{
C(\rr-\rr',\tau)
\notag\\
&\times
\left[
\hat n_\psi(\rr),
\hat n_\psi(\rr',-\tau)\,
\rho_\psi(t)
\right]
+\mathrm{h.c.}
\Big\}.
\label{eq:MES}
\end{align}

\par
We now derive the equation of motion of the condensate,

\begin{equation}
\frac{d}{dt}
\langle
\hat\psi(\mathbf{x})
\rangle
=
\Tr
\left[
\hat\psi(\mathbf{x})
\,
\frac{\partial \rho_\psi(t)}{\partial t}
\right].
\label{eq:EM_psi}
\end{equation}
%%continue from here
Since the condensate is highly occupied and coherent, we can replace products of operators with their expectation values. Equation \eqref{eq:MES} has two parts --- the condensate commutator and condensate-reservoir commutator--- and we evaluate them in turn. Using the notation $\psi(\rr,t)=\langle\hat\psi(\rr)\rangle$ and the bosonic commutations, the first term then gives the familiar G-P equation. The second term is computed using the same tools. Substituting Eq.~\eqref{eq:MES} into Eq.~\eqref{eq:EM_psi} yields two trace terms. Using
\( [\hat\psi(\mathbf{x}),\hat n_\psi(\rr)] = \delta^{(D)}(\mathbf{x}-\rr)\,\hat\psi(\mathbf{x}) \), these reduce to
\begin{align*}
&\Tr
\big[
\hat\psi(\mathbf{x})
\left[
\hat n_\psi(\rr),
\hat n_\psi(\rr',-\tau)\rho_\psi
\right]
\big]
\notag\\
&\qquad\qquad=
\delta^{(D)}(\mathbf{x}-\rr)
\left<
\hat\psi(\mathbf{x})
\hat n_\psi(\rr',-\tau)
\right>,
\\[2pt]
&\Tr
\big[
\hat\psi(\mathbf{x})
\left[
\rho_\psi
\hat n_\psi(\rr',-\tau),
\hat n_\psi(\rr)
\right]
\big]
\notag\\
&\qquad\qquad=
-
\delta^{(D)}(\mathbf{x}-\rr)
\left<
\hat n_\psi(\rr',-\tau)
\hat\psi(\mathbf{x})
\right>.
\end{align*}

Applying the coherent-field factorization described above, identifying the expectation value of the backward-evolved density operator with the condensate density at the delayed time to the present order in the system–reservoir interaction, and evaluating the $d^{D}\rr$ integral in Eq.~\eqref{eq:MES} using the $\delta$-function yields the full equation of motion for the condensate

\begin{equation}
i\hbar\,\frac{\partial\psi(\mathbf{x},t)}{\partial t}
=
\left[-\frac{\hbar^2\nabla^2}{2m_\psi}+V_{\rm ext}+U|\psi|^2
+V_{\rm drag}(\mathbf{x},t)\right]\psi,
\label{eq:full}
\end{equation}
The reservoir enters as the potential in the G-P equation $V_{\rm drag}$ given by
\begin{align}
V_{\rm drag}(\mathbf{x},t)
={}&
\frac{2g^2}{\hbar}
\int_0^\infty\!d\tau\!
\int d^{D}\rr'\,
\operatorname{Im}C(\mathbf{x}-\rr',\tau)
\notag\\
&\times
\left|\psi(\rr',t-\tau)\right|^2 .
\label{eq:vdrag}
\end{align}

where we have used the fact $C-C^*=2i\,\mathrm{Im}\,C$. Within the coherent-field approximation, the influence of the reservoir is contained in the single real, history-dependent potential $V_{\rm drag}$, determined by the reservoir density correlator and the microscopic interaction strength $g$, without introducing an additional phenomenological drag coefficient.

Here ``history-dependent'' refers to the fact that the drag potential at time $t$ depends on the condensate density
at a retarded time $t-\tau$. Within the Born--Markov approximation, the reservoir correlation function $C(\mathbf q,\tau)$ decays over a correlation time $\tau_c$ that is short compared with the characteristic evolution time of the condensate, so only the condensate evolution over this short interval contributes to the integral. The retarded density may therefore be expanded systematically in powers of $\tau$, as done below in Eq.~\eqref{eq:comoving_taylor}, yielding an effective potential that is local in time to the order considered.

The influence of the reservoir enters through its density correlation function $C(\mathbf q,\tau)$; from this point onward, we specialize to a fermionic reservoir and explicitly evaluate this correlator. The remaining task is to evaluate $C(\rr-\rr',\tau)$ for the drifting fermionic reservoir.

\begin{equation*}
C(\rr-\rr',\tau)
=
\int\frac{d^{D}\q}{(2\pi)^{D}}\,
e^{i\q\cdot(\rr-\rr')}\,
C(\q,\tau),
\end{equation*}
with
\[
C(\q,\tau)
\equiv
\frac{1}{L^{D}}
\left\langle
\delta \hat n_\q(\tau)\,
\delta \hat n_{-\q}(0)
\right\rangle ,
\]
where $L^{D}$ is the volume of the system. The reservoir density operator in momentum space is given by
\[
\delta\hat n_{\q}(t)
=
\sum_{\kk}
\hat c^\dagger_{\kk}(t)\;
\hat c_{\kk+\q}(t),
\]
where $\hat c_{\kk}(t) = \hat c_{\kk} e^{-iE_{\kk}t/\hbar}$. Substituting the density operators,
\begin{align*}
C(\q,\tau)
&=
\frac{1}{L^{D}}
\left\langle
\delta\hat n_{\q}(\tau)\,
\delta\hat n_{-\q}(0)
\right\rangle
\notag\\
&=
\frac{1}{L^{D}}
\sum_{\kk,\kk'}
e^{-i(E_{\kk+\q}-E_{\kk})\tau/\hbar}
\left\langle
\hat c_{\kk}^{\dagger}
\hat c_{\kk+\q}
\hat c_{\kk'}^{\dagger}
\hat c_{\kk'-\q}
\right\rangle.
\end{align*}
For this term to be nonzero, creation operators must undo the destruction operators, which gives
\begin{equation*}
C(\q,\tau)
=
\frac{1}{L^{D}}
\sum_{\kk}
f(\kk)
\left[1-f(\kk+\q)\right]
e^{-i(E_{\kk+\q}-E_{\kk})\tau/\hbar},
\end{equation*}
which in the continuum limit becomes
\begin{equation}
C(\q,\tau)
=
\int\frac{d^{D}\kk}{(2\pi)^{D}}\,
f(\kk)
\left[1-f(\kk+\q)\right]
e^{-i(E_{\kk+\q}-E_{\kk})\tau/\hbar}.
\label{eq:Cqcont}
\end{equation}
So far, the occupation number of the reservoir has been left in an arbitrary state $f(\kk)$. The current can be introduced in the occupation number using a drifted distribution $f(\kk-\kk_d)$, where $\hbar\kk_d=m_\phi\mathbf v_d $. Substituting the drifted distribution into \eqref{eq:Cqcont}
and changing the integration variable $\kk\to\kk+\kk_d$, the occupation factors return
to their rest form, while the energy difference in the exponential factor picks up one extra term,
\begin{align*}
E_{\kk+\kk_d+\q}-E_{\kk+\kk_d}
&=\frac{\hbar^2}{2m_\phi}\big[(\kk+\q)^2-\kk^2+2\kk_d\cdot\q\big]
\notag\\
&=\big(E_{\kk+\q}-E_{\kk}\big)+\hbar\,\q\cdot\mathbf v_d ,
\end{align*}
which factors out of the integral as a pure phase:
\begin{equation*}
\;C_{v_d}(\q,\tau)=e^{-i\,\q\cdot\mathbf v_d\,\tau}\;C_0(\q,\tau).\;
\end{equation*}
where $C_0(\q,\tau)$ is the reservoir density correlation function in the absence of drift. Therefore, the effect of the current on the bath enters as a phase factor $e^{-i\,\q\cdot\mathbf v_d\,\tau}$. Thus, we have
\begin{equation}
C(\rr-\rr',\tau)
=
\int\frac{d^{D}\q}{(2\pi)^{D}}
e^{i\q\cdot(\rr-\rr')}
e^{-i\,\q\cdot\mathbf v_d\,\tau}\;C_0(\q,\tau)
\label{eq:Crfinal}
\end{equation}
Substituting Eq.~\eqref{eq:Crfinal} into Eq.~\eqref{eq:vdrag}, we obtain
\begin{align*}
V_{\rm drag}(\mathbf{x},t)
={}&
\frac{2g^2}{\hbar}
\int_0^\infty \!d\tau\!
\int \!d^{D}\rr'
\!\int\! \frac{d^{D}\q}{(2\pi)^{D}}
\left|\psi(\mathbf{r}',t-\tau)\right|^2
\notag\\
&\times
\operatorname{Im}
\left[
e^{i\mathbf{q}\cdot(\mathbf{x}-\mathbf{r}')}
e^{-i\mathbf{q}\cdot\mathbf{v}_d\tau}
C_0(\mathbf{q},\tau)
\right].
\end{align*}
Defining the shifted coordinate $\mathbf r''=\mathbf r'+\mathbf v_d\tau$ gives
\begin{align}
V_{\rm drag}(\mathbf{x},t)
={}&
\frac{2g^2}{\hbar}
\int_0^\infty \!d\tau\!
\int \!d^{D}\rr''
\!\int\! \frac{d^{D}\q}{(2\pi)^{D}}
\notag\\
&\times
\left|
\psi\!\left(
\mathbf r''-\mathbf v_d\tau,
t-\tau
\right)
\right|^2
\notag\\
&\times
\operatorname{Im}
\left[
e^{i\mathbf q\cdot(\mathbf x-\mathbf r'')}
C_0(\mathbf q,\tau)
\right].
\label{eq:vdrag_shifted}
\end{align}
During the timescale of the reservoir, we can take the condensate density to be slowly varying in both space and time, allowing us to expand the condensate density to first order
\begin{align}
n_\psi(\mathbf r''-\mathbf v_d\tau,\,t-\tau)
={}&
n_\psi(\mathbf r'',t)
\notag\\
&-\tau\big[\partial_t+\mathbf v_d\cdot\nabla\big]\,n_\psi(\mathbf r'',t)
\notag\\
&+\mathcal O(\tau^2).
\label{eq:comoving_taylor}
\end{align}
Substituting \eqref{eq:comoving_taylor} into \eqref{eq:vdrag_shifted}, gives
\begin{align}
V_{\rm drag}(\mathbf{x},t)
={}&
\frac{2g^2}{\hbar}
\int d^{D}\rr''\!\int\!\frac{d^{D}\q}{(2\pi)^{D}}\,
\notag\\
&\times\Big\{
\operatorname{Im}\!\big[e^{i\q\cdot(\mathbf x-\mathbf r'')}K_0(\q)\big]\,
n_\psi(\mathbf r'',t)
\notag\\
&\quad-\operatorname{Im}\!\big[e^{i\q\cdot(\mathbf x-\mathbf r'')}K_1(\q)\big]
\notag\\
&\qquad\times
\big[\partial_t+\mathbf v_d\cdot\nabla\big]\,n_\psi(\mathbf r'',t)
\Big\},
\label{eq:vdrag_moments}
\end{align}
where
\begin{equation}
K_m(\q)\equiv\int_0^\infty d\tau\;\tau^m\,C_0(\q,\tau).
\qquad m=0,1,
\label{eq:moments}
\end{equation}
The $K_0$ term is  proportional to the density, leading to a static term, which is independent of reservoir velocity. The drag arises entirely from the $K_1$ contribution, which is proportional to $(\partial_t+\mathbf v_d\!\cdot\nabla)n_\psi$.

For the case of a number-conserving system, we can rewrite the term $(\partial_t+\mathbf v_d\!\cdot\nabla)n_\psi$ by using the fact
\begin{align}
\frac{\partial n_\psi}{\partial t}
&=
\psi^*\frac{\partial\psi}{\partial t}
+
\psi\frac{\partial\psi^*}{\partial t}=
2\,\operatorname{Re}
\left(
\psi^*\frac{\partial\psi}{\partial t}
\right).
\label{eq:density_derivative}
\end{align}
Using Eq.~\eqref{eq:full} gives
\begin{align*}
\frac{\partial n_\psi}{\partial t}
={}&
2\,\operatorname{Re}
\Bigg[
\frac{i\hbar}{2m_\psi}\psi^*\nabla^2\psi
\notag\\
&-
\frac{i}{\hbar}
\left(
V_{\rm ext}
+
U|\psi|^2
+
V_{\rm drag}
\right)
|\psi|^2
\Bigg].
\end{align*}
Since $V_{\rm ext}$, $U|\psi|^2$, and $V_{\rm drag}$ are all real and the second term is purely
imaginary, the continuity equation becomes

\begin{align*}
\frac{\partial n_\psi}{\partial t}
&=
\frac{i\hbar}{2m_\psi}
\left(
\psi^*\nabla^2\psi
-
\psi\nabla^2\psi^*
\right)
\nonumber=
-\nabla\cdot\mathbf j_\psi.
\end{align*}
Thus, although $V_{\rm drag}$ can redistribute the condensate momentum, the drag potential does not appear as a source or loss term in the density dynamics. In the next section, this continuity equation is modified by the inclusion of incoherent pumping and decay, which introduce gain and loss terms appropriate for a driven--dissipative condensate.

Using the continuity equation, Eq.~\eqref{eq:vdrag_moments} becomes
\begin{equation*}
\left(
\partial_t+\mathbf v_d\cdot\nabla
\right)n_\psi
=
-\nabla\cdot\mathbf j_\psi
+
\mathbf v_d\cdot\nabla n_\psi.
\end{equation*}
Thus,
\begin{align}
V_{\rm drag}(\mathbf{x},t)
={}&
\frac{2g^2}{\hbar}
\int d^{D}\rr''\!\int\!\frac{d^{D}\q}{(2\pi)^{D}}
\notag\\
&\times\Bigg\{
\operatorname{Im}
\!\left[
e^{i\q\cdot(\mathbf x-\mathbf r'')}
K_0(\q)
\right]
n_\psi(\mathbf r'',t)
\notag\\
&\quad+
\operatorname{Im}
\!\left[
e^{i\q\cdot(\mathbf x-\mathbf r'')}
K_1(\q)
\right]
\notag\\
&\qquad\times
\big[
\nabla\cdot\mathbf j_\psi(\mathbf r'',t)
-
\mathbf v_d\cdot\nabla n_\psi(\mathbf r'',t)
\big]
\Bigg\}.
\label{eq:vdrag_moments2}
\end{align}

Equation~\eqref{eq:vdrag_moments2} shows that the drag is fundamentally a dynamical effect, which depends on the relative current of the condensate to the reservoir. Since the interaction of the condensate with the reservoir enters only through a real potential, it only redistributes condensate particles without changing their total number.

A more physically transparent form is obtained by integrating Eq.~\eqref{eq:vdrag_moments2} by parts in $\mathbf r''$ (Appendix~\ref{app:altform}). The resulting surface term vanishes for periodic boundaries and, more generally, whenever the condensate density and current vanish at the boundary of the system --- as they do for a spatially localized condensate or at hard walls, where $\psi=0$ forces $n_\psi=\mathbf j_\psi=0$. Under this assumption, the drag potential becomes

\begin{align*}
V_{\rm drag}(\mathbf x,t)
={}&
\frac{2g^2}{\hbar}
\int d^{D}\rr''
\int\!\frac{d^{D}\q}{(2\pi)^{D}}
\Bigg\{
\notag\\
&
\operatorname{Im}\!\left[
e^{i\q\cdot(\mathbf x-\mathbf r'')}
K_0(\q)
\right]
n_\psi(\mathbf r'',t)
\notag\\
&
+\,
\q\cdot
\Big[
n_\psi(\mathbf r'',t)
\big(
\mathbf v_\psi(\mathbf r'',t)-\mathbf v_d
\big)
\Big]
\notag\\
&
\times
\operatorname{Re}\!\left[
e^{i\q\cdot(\mathbf x-\mathbf r'')}
K_1(\q)
\right]
\Bigg\},
\end{align*}
where $\mathbf v_\psi=\mathbf j_\psi/n_\psi$ is the local condensate velocity. This form makes it clear that the second term depends on the relative velocity through $\mathbf j_\psi-\mathbf v_d n_\psi=n_\psi(\mathbf v_\psi-\mathbf v_d)$, making the drag vanish wherever $\mathbf v_\psi=\mathbf v_d$, leaving only the screening contribution.

\section*{Fluctuation--dissipation relation}

$K_0$ and $K_1$ obey a simple identity for a thermal reservoir in the Maxwell--Boltzmann regime. The $K_0$ leads to a static screening potential, independent of the condensate--reservoir velocity, while the $K_1$ term leads to the drag effect and is proportional to the relative velocity.

Up to this point, the fermionic occupation distribution has been left general. We now specialize to the dilute, nondegenerate regime relevant to our application, where the reservoir is described by a Maxwell--Boltzmann distribution. In this low-density limit, $f(k)\left[1-f(k+q)\right]\simeq f(k)$, and Eq.~\eqref{eq:Cqcont} becomes

\begin{equation}
C_0(\mathbf q,\tau)
=
n_{\phi}
\exp\!\left(
-\frac{i\hbar q^2\tau}{2m_\phi}
\right)
\exp\!\left(
-\frac{k_BT\,q^2}{2m_\phi}\tau^2
\right),
\label{eq:C_MB_general}
\end{equation}
Differentiating Eq.~\eqref{eq:C_MB_general} gives
\begin{equation*}
\frac{\partial C_0}{\partial \tau}
=
-
\left(
\frac{i\hbar q^2}{2m_\phi}
+
q^2v_T^2\tau
\right)
C_0,
\end{equation*}
where $v_T^2=k_BT/m_\phi$. Integrating over $\tau$ from $0$ to $\infty$ yields the exact recursion relation
\begin{equation*}
q^2v_T^2K_1(\mathbf q)
=
n_{\phi}
-
\frac{i\hbar q^2}{2m_\phi}K_0(\mathbf q),
\end{equation*}
Taking the imaginary part gives
\begin{equation*}
\operatorname{Im}K_1(\mathbf q)
=
-\frac{\hbar}{2k_BT}
\operatorname{Re}K_0(\mathbf q).
\end{equation*}
Since $C_0(\mathbf q,-\tau)=C_0(\mathbf q,\tau)^*$, Eq.~\eqref{eq:moments} implies
\begin{equation*}
2\operatorname{Re}K_0(\mathbf q)
=
\int_{-\infty}^{\infty}
d\tau\,
C_0(\mathbf q,\tau),
\end{equation*}
so that
\begin{equation}
-\operatorname{Im}K_1(\mathbf q)
=
\frac{\hbar}{4k_BT}
\int_{-\infty}^{\infty}
d\tau\,
C_0(\mathbf q,\tau).
\label{eq:FDT_final}
\end{equation}

Equation~\eqref{eq:FDT_final} has the form of a classical zero-frequency fluctuation--dissipation relation \cite{kubo1966fluctuation,callen1951irreversibility}: the dissipative drag is directly proportional to the density fluctuations of the reservoir. The drag kernel $\operatorname{Im}K_1$ is that weight multiplied by $-\hbar/4k_BT$ --- friction and fluctuations share one origin. For a Maxwell--Boltzmann thermal reservoir, the time-integrated density correlation is nonnegative, and Eq.~\eqref{eq:FDT_final} therefore implies that $-\operatorname{Im}K_1(q)\geq 0$ for every $q$. This sign ensures that the drag opposes the relative motion. The reservoir therefore transfers momentum so as to drive the condensate toward co-motion with the reservoir, $\mathbf v_\psi=\mathbf v_d$.

\section*{Effective drag force}

For a number-conserving condensate, because $V_{\rm drag}$ enters the equation
of motion as a real potential, the net force it exerts on the condensate is given by
\cite{Astrakharchik2004,wouters2010superfluidity},
\begin{equation}
\mathbf F
=
-\int d^{D}r\,
n_\psi(\rr,t)\,\nabla V_{\rm drag}(\rr,t).
\label{eq:force_def}
\end{equation}
Writing $V_{\rm drag}=V^{(0)}+V^{(1)}$, the force similarly separates as
$\mathbf F=\mathbf F^{(0)}+\mathbf F^{(1)}$, where $V^{(0)}$ is the static
screening contribution proportional to $K_0$, while $V^{(1)}$ is the
velocity-dependent contribution proportional to $K_1$. Substituting the screening part of $V_{\rm drag}$ into
Eq.~\eqref{eq:force_def} and evaluating the spatial integrals gives
\begin{equation*}
\mathbf F^{(0)}
=
-\frac{2g^2}{\hbar}
\int\frac{d^{D}q}{(2\pi)^{D}}\,
\q\,
\operatorname{Re}K_0(q)\,
|n_\q|^2.
\end{equation*}

where $n_\q$ is given by
\begin{equation*}
n_\q
=
\int d^{D}r\,
n_\psi(\rr)e^{-i\q\cdot\rr},
\end{equation*}

For an isotropic system, the integrand is odd under $\q\rightarrow-\q$, because $\q$ is odd while
$\operatorname{Re}K_0(q)$ and $|n_\q|^2$ are even. Therefore, $\mathbf F^{(0)}=0$.  Thus, although the $K_0$ term modifies the condensate potential, it exerts no net force on the condensate.

The velocity-dependent part gives
\begin{equation}
\mathbf F^{(1)}
=
\frac{2g^2}{\hbar}
\int\frac{d^{D}q}{(2\pi)^{D}}\,
\q\,
\operatorname{Im}
\left[
K_1(q)
\left(
\q\cdot\mathbf J_\q
\right)
n_\q^{*}
\right],
\label{eq:F1exact}
\end{equation}
where $J_q$ is the Fourier-transform of $\mathbf J(\rr) = n_\psi(\rr) \left[ \mathbf v_\psi(\rr)-\vd \right]$
\begin{equation*}
\mathbf J_\q
=
\int d^{D}r\,
\mathbf J(\rr)e^{-i\q\cdot\rr},
\end{equation*}
Equation~\eqref{eq:F1exact} is exact for an arbitrary condensate density and
current profile. To obtain a simple friction law, we assume that the relative velocity $\mathbf u=\mathbf v_\psi-\vd$ is approximately uniform across the condensate. In this case,
$\mathbf J(\rr)=\mathbf u\,n_\psi(\rr)$ and hence
$\mathbf J_\q=\mathbf u\,n_\q$. Equation~\eqref{eq:F1exact} then becomes
\begin{equation*}
\mathbf F^{(1)}
=
\frac{2g^2}{\hbar}
\int\frac{d^{D}q}{(2\pi)^{D}}\,
\q\,
\left(
\q\cdot\mathbf u
\right)
|n_\q|^2\,
\operatorname{Im}K_1(q).
\end{equation*}
For an isotropic condensate, $|n_\q|^2$ depends only on $|q|$, and we obtain
\begin{equation*}
\mathbf F
=
-\Gamma_d
\left(
\mathbf v_\psi-\vd
\right),
\end{equation*}
where

\begin{align*}
\Gamma_d
&=
-\frac{2g^2}{\hbar D}
\int\frac{d^{D}q}{(2\pi)^{D}}\,
q^2\,|n_\q|^2\,
\operatorname{Im}K_1(q)
\notag\\
&=
\frac{g^2}{2Dk_BT}
\int\frac{d^{D}q}{(2\pi)^{D}}\,
q^2\,|n_\q|^2
\int_{-\infty}^{\infty}d\tau\,
C_0(q,\tau),
\end{align*}
where in the second line we used the fluctuation--dissipation relation given by Eq.~\eqref{eq:FDT_final}, which
implies that $-\operatorname{Im}K_1(q)\geq0$, and therefore $\Gamma_d>0$. Therefore, the drag force opposes the relative motion and transfers momentum to drive the condensate toward co-motion with the reservoir.

For a Maxwell--Boltzmann reservoir, the correlator $C_0(q,\tau)$ is given by
Eq.~\eqref{eq:C_MB_general}, and the friction coefficient becomes
\begin{align*}
\Gamma_d=
\frac{\sqrt{\pi}\,g^2n_\phi\ell_T}
{\hbar Dv_T^2}
\int\frac{d^{D}q}{(2\pi)^{D}}\,
|q|\,|n_\q|^2
e^{-q^2\ell_T^2/4}.
\end{align*}
Within the coherent-field approximation considered here, the dependence of the friction on the condensate profile is entirely contained in its density form factor. Since the integral vanishes at $q=0$, a spatially uniform condensate does not experience drag through this mechanism; momentum exchange with the reservoir arises from finite-$q$ density variations. We note that If the relative condensate--reservoir velocity exceeds the Landau critical velocity, additional dissipative processes associated with the creation of excitations can produce drag even in a uniform system. Such a regime can, in principle, be studied by linearizing the Gross--Pitaevskii equation about the uniform solution and examining the resulting Bogolyubov excitation spectrum.

\section*{Numerical Simulations}

\begin{figure}
\includegraphics[width=1\columnwidth]{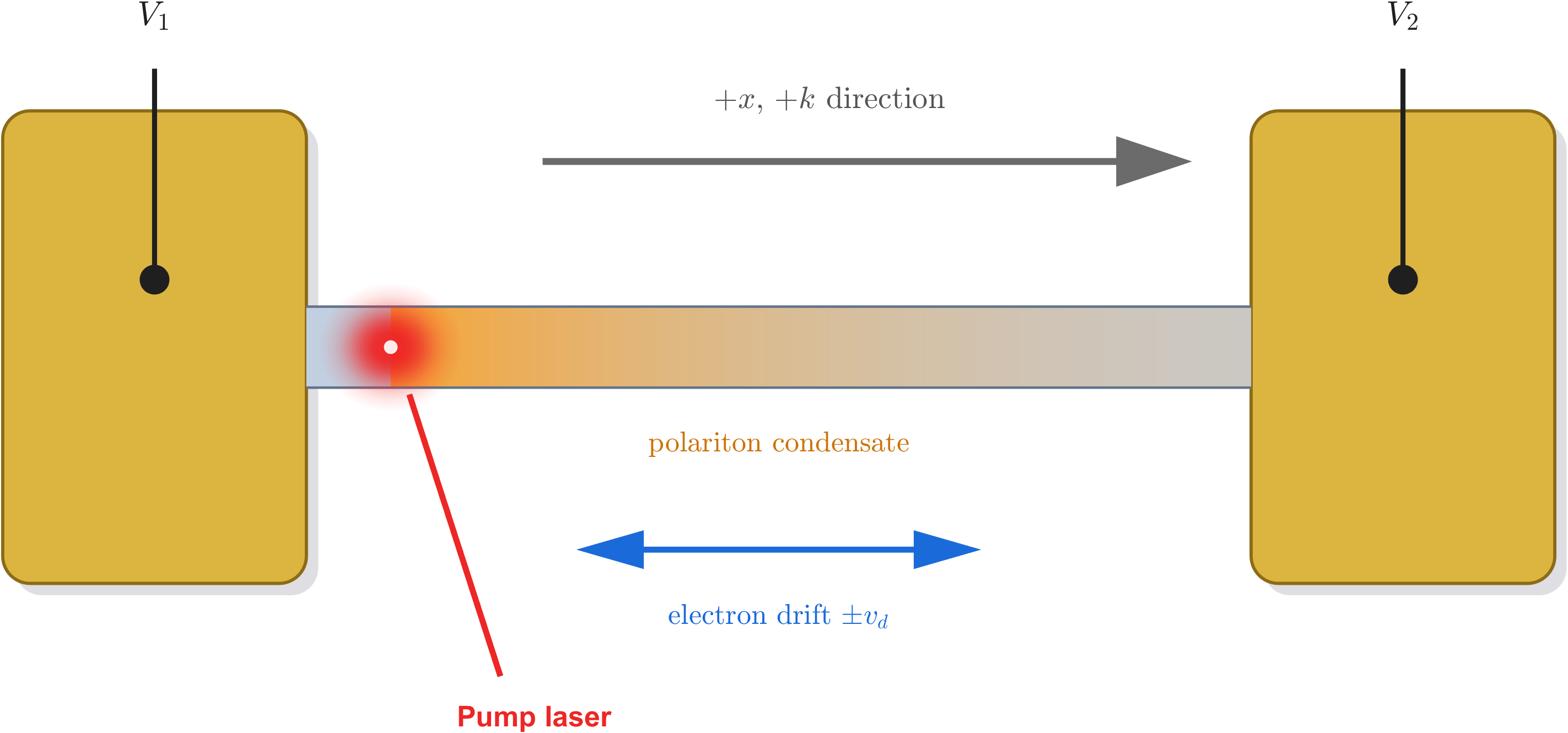}
\centering
\caption{Schematic of the simulated polaritons in a one-dimensional wire. A non-resonant pump laser shown in red is focused near one end of the wire generating a polariton condensate at the pump spot and propagating ballistically along the wire (orange). A voltage bias is applied between the contacts driving a current of electrons through the wire; depending on the sign of the voltage, the electron drift velocity $\pm v_d$ can be positive or negative, so that the electron flow can be made either co-propagating or counter-propagating with the condensate. The condensate flow direction defines the positive $x$ (equivalently $+k$) axis.}
\label{fig:wire}
%%%%%%%%%%%%%%%%%%%%%%%%%%%
\end{figure}

As an application of the theory, we numerically model a one-dimensional polariton wire using parameters and geometry similar to those of the experimental system reported in~\cite{Myers2025}, shown schematically in Fig.~\ref{fig:wire}. The wire is assumed to be electron doped so that an electric current can be driven through it by an applied voltage. The resulting reservoir drift velocity $\mathbf{v}_d$ induced by the voltage difference serves as the tunable control parameter for the condensate--reservoir drag.

The system is described by two coupled equations: an open-dissipative G-P equation for the polariton condensate and a kinetic equation for the incoherent exciton reservoir \cite{Wouters2007PRL,Wouters2008SM,fontaine2022kardar,alnatah2024critical}. The microscopic derivation above determines the electron-induced drag potential $V_{\rm drag}$. The pumping, decay, and incoherent exciton-reservoir dynamics of the polariton system are not derived from the electron bath. For the application considered here, we therefore insert the microscopically derived drag potential into the standard phenomenological open-dissipative G-P model of a polariton condensate.

The condensate equation of motion obeys
\begin{align}
i\hbar\frac{\partial\psi(x,t)}{\partial t}
={}&
\Bigg[
-\frac{\hbar^2}{2m_\psi}\frac{\partial^2}{\partial x^2}
+U\left|\psi(x,t)\right|^2
\nonumber\\
&+U_R n_R(x,t)
+V_{\rm drag}(x,t)
\nonumber\\
&+\frac{i\hbar}{2}
\left(
R n_R(x,t)-\gamma_c
\right)
\Bigg]\psi(x,t),
\label{eq:driven_GPE_reservoir}
\end{align}
while the exciton reservoir density satisfies
\begin{equation}
\frac{\partial n_R(x,t)}{\partial t}
= P(x,t)
-
\left[
\gamma_R
+
R\left|\psi(x,t)\right|^2
\right]n_R(x,t).
\label{eq:reservoir_dynamics_numerical}
\end{equation}
Here, $m_\psi$ is the polariton effective mass, $U$ is the polariton--polariton interaction strength, $U_R$ is the repulsive polariton--excitonic reservoir interaction strength, $R$ is the stimulated scattering rate from the reservoir into the condensate, $\gamma_c$ and $\gamma_R$ are the condensate and reservoir decay rates, respectively, and $P(x,t)$ is the nonresonant pump. We take $\gamma_c=1/(270~\mathrm{ps})$, consistent with the long-lifetime polaritons used in the experiment of Ref.~\cite{Myers2025}. Such long-lifetime polaritons \cite{steger2015slow} have been shown to reach thermal equilibrium in Refs.~\cite{sun2017bose,alnatah2024coherence,alnatah2024bose,alnatah2025strong}. The electron-induced drag is included through the potential $V_{\rm drag}(x,t)$.

We adiabatically eliminate the reservoir by setting $\partial n_R/\partial t =0$ and substituting the steady-state reservoir density $n_R(x,t)$ into Eq.~\eqref{eq:driven_GPE_reservoir}, which gives \cite{bobrovska2015adiabatic,carusotto2013quantum,comaron2018dynamical}
\begin{align}
i\hbar\frac{\partial\psi(x,t)}{\partial t}
={}&
\Bigg[
-\frac{\hbar^2}{2m_\psi}\frac{\partial^2}{\partial x^2}
+U|\psi(x,t)|^2
\nonumber\\
&+V_{\rm drag}(x,t)
+
\frac{U_R P(x,t)}
{\gamma_R+R|\psi(x,t)|^2}
\nonumber\\
&+
\frac{i\hbar}{2}
\left(
\frac{R P(x,t)}
{\gamma_R+R|\psi(x,t)|^2}
-\gamma_c
\right)
\Bigg]\psi(x,t).
\label{eq:effective_driven_GPE}
\end{align}

The polariton-electron interaction $g$ enters through the drag potential. The direct Coulomb interaction between an electron and a polariton is negligible since the excitonic component of the polariton is electrically neutral. The dominant contribution instead comes from exchange between the free electron and the electron within the exciton, resulting in an effective repulsive electron--polariton interaction. The interaction $g$ can be treated as a contact interaction since its momentum dependence, $g(k)$, varies over a characteristic scale of order $1/a_B$, where $a_B$ is the exciton Bohr radius \cite{ramon2002scattering,li2021theory}. Since the polariton thermal wavelength is $\lambda_p  \sim  1~\mu\mathrm{m}$, which is much larger than $a_B$, the polaritons probe only the nearly constant, small-$k$ region of $g(k)$ and can be treated as a constant.

\begin{figure}
\includegraphics[width=0.8\columnwidth]{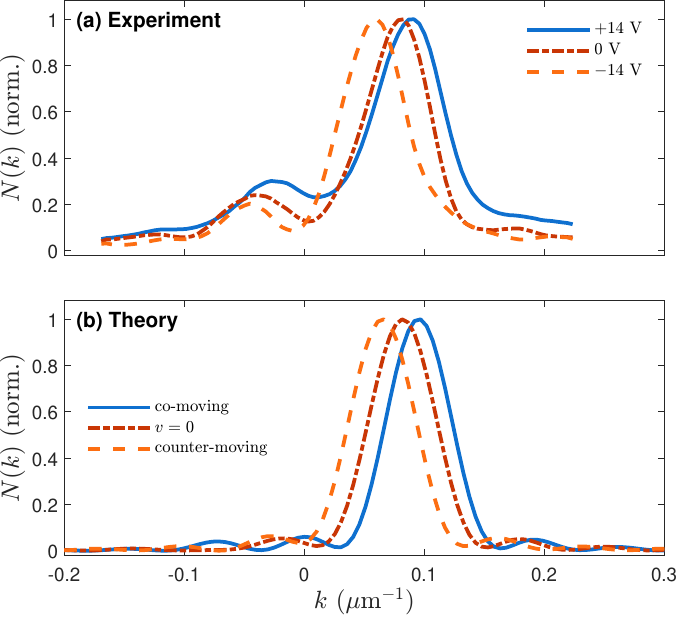}
\centering
\caption{Momentum-space drag of the polariton condensate. (a) Measured PL intensity vs.\ in-plane momentum $k$ (reproduced from Ref.~\cite{Myers2025}) at $0$ V ($v=0$), $+14$ V ($+115.4$ $\mu$A, co-moving), and $-14$ V ($-143.4$ $\mu$A, counter-moving). (b) Simulated momentum distribution $N(k)$ from the driven–dissipative Gross–Pitaevskii equation with the microscopic drag potential, same three cases ($v_d=0,\pm0.14$ $\mu$m/ps). In both, the forward peak shifts with the electron current (co $\to$ higher $k$, counter $\to$ lower $k$).}
\label{fig:N_k}
\end{figure}

We evaluate the rest-frame correlator $C_0$ for a dilute electron gas in thermal equilibrium, such that $f(k)\left[1-f(k+q)\right]\simeq f(k)$, with the occupation described by the Maxwell--Boltzmann distribution.
\begin{align*}
C_0(q,\tau)
&=
\int\frac{dk}{2\pi}\,
f(k)\,
e^{-i(E_{k+q}-E_k)\tau/\hbar}
\nonumber\\
&=
\int\frac{dk}{2\pi}\,
e^{\mu/(k_B T)}
e^{-\hbar^2 k^2/(2m_\phi k_B T)}
\nonumber\\
&\hspace{4.2em}\times
e^{-i\hbar(2kq+q^2)\tau/(2m_\phi)}.
\end{align*}
Evaluating the integral gives
\begin{equation*}
C_0(q,\tau)
=
n_e
\exp\!\left(-\frac{i\hbar q^2\tau}{2m_\phi}\right)
\exp\!\left(-\frac{q^2 k_B T\,\tau^2}{2m_\phi}\right),
\end{equation*}
where $n_e=e^{\mu/k_B T}\sqrt{m_\phi k_B T/2\pi\hbar^2}$ is the one-dimensional electron density. Defining $v_T=\sqrt{k_B T/m_\phi}$, \, $\ell_T=\hbar/\sqrt{2m_\phi k_B T}$, the Gaussian integrals become
\begin{align*}
K_0(q)&=\frac{n_e}{|q|v_T}
\left[\sqrt{\tfrac{\pi}{2}}\,e^{-q^{2}\ell_T^{2}/4}
\;-\;i\sqrt{2}\,D\!\left(\tfrac{|q|\ell_T}{2}\right)\right],
\\[4pt]
K_1(q)&=\frac{n_e}{q^{2}v_T^{2}}
\Big[1-|q|\ell_T\,D\!\left(\tfrac{|q|\ell_T}{2}\right)
\nonumber\\
&\hspace{4.6em}
-\;i\,\frac{\sqrt{\pi}}{2}\,|q|\ell_T\,e^{-q^{2}\ell_T^{2}/4}\Big],
\end{align*}
where $D(s)=e^{-s^{2}}\!\int_0^{s}e^{t^{2}}dt$ is the Dawson function. Both $K_0$ and $K_1$ are even in $q$, so the terms odd in $q$ vanish after integration. Substituting the expressions for $K_0$ and $K_1$ into Eq.~\eqref{eq:vdrag_moments} and evaluating the  momentum integrals gives
\begin{align}
V_{\rm drag}(x,t)
={}&
-\,\frac{g^{2}n_e}{\hbar \,v_T}\sqrt{\frac{\pi}{2}}
\int dx''\;\operatorname{erfc}\!\left(\frac{|x-x''|}{\ell_T}\right)
\nonumber\\
&\hspace{2.6em}\times\, n_\psi(x'',t)
\nonumber\\[2pt]
&+\,\frac{g^{2}n_e\,\ell_T}{\sqrt{\pi}\,\hbar \,v_T^{2}}
\int dx''\,
\bigg[\ln\frac{L}{\pi\ell_T}-\frac{\gamma_E}{2}
\nonumber\\
&\hspace{5.2em}
-2\!\!\int_0^{|x-x''|/\ell_T}\!\!\!\!D(s)\,ds\bigg]
\nonumber\\
&\hspace{2.6em}\times
\big[\partial_t+v_d\,\partial_{x''}\big] n_\psi(x'',t),
\label{eq:vdrag_1d}
\end{align}
Here, $L$ denotes the finite length of the polariton wire and therefore sets
the infrared cutoff of the one-dimensional momentum integral.

Equation~\eqref{eq:vdrag_1d} can be simplified by using the fact $\partial_t n_\psi=2\operatorname{Re}(\psi^*\partial_t\psi)$, which through Eq.~\eqref{eq:effective_driven_GPE} gives
\begin{equation}
\frac{\partial n_\psi}{\partial t}
=
-\partial_x j_\psi
+
\left(
\frac{R P}{\gamma_R+R|\psi|^2}
-\gamma_c
\right)
n_\psi ,
\label{eq:density_rate_driven}
\end{equation}
with $j_\psi=(\hbar/m_\psi)\operatorname{Im}(\psi^*\partial_x\psi)$. The continuity
equation thus includes the gain and loss terms, and the drag kernel in
Eq.~\eqref{eq:vdrag_1d} acts on
\begin{align}
\big[\partial_t+v_d\,\partial_{x''}\big]n_\psi
={}&
-\partial_{x''} j_\psi
+v_d\,\partial_{x''} n_\psi
\notag\\
&+
\left(
\frac{R P}{\gamma_R+R|\psi|^2}
-\gamma_c
\right)
n_\psi .
\label{eq:comoving_driven}
\end{align}
Because $V_{\rm drag}$ is real, it does not contribute to Eq.~\eqref{eq:density_rate_driven}. Substituting Eq.~\eqref{eq:comoving_driven} into Eq.~\eqref{eq:vdrag_1d} and integrating by parts gives (Appendix~\ref{app:relcurrent}) 
\begin{align}
V_{\rm drag}(x,t)
={}&
-\frac{g^{2}n_e}{\hbar\,v_T}\sqrt{\frac{\pi}{2}}
\int dx''\,
\operatorname{erfc}\!\left(\frac{|x-x''|}{\ell_T}\right)
\notag\\
&\hspace{3.2em}\times n_\psi(x'',t)
\notag\\[2pt]
&+\frac{2g^{2}n_e}{\sqrt{\pi}\,\hbar\,v_T^{2}}
\int dx''\,
D\!\left(\frac{x-x''}{\ell_T}\right)
\notag\\
&\hspace{3.2em}\times
n_\psi(x'',t)\,\big[\,v_\psi(x'',t)-v_d\,\big]
\notag\\[2pt]
&+\frac{g^{2}n_e\,\ell_T}{\sqrt{\pi}\,\hbar\,v_T^{2}}
\int dx''\,
\mathcal F(x-x'')
\notag\\
&\hspace{3.2em}\times
\Gamma(x'',t)\,n_\psi(x'',t),
\label{eq:vdrag_relative_1d}
\end{align}
where $v_\psi=j_\psi/n_\psi$ is the local condensate velocity,
$\Gamma(x,t)=RP/(\gamma_R+Rn_\psi)-\gamma_c$ contains the local gain and loss, and
$\mathcal F(x-x'')=\ln(L/\pi\ell_T)-\gamma_E/2-2\int_0^{|x-x''|/\ell_T}\!D(s)\,ds$. The
first two terms are conservative: a static screening shift set by the density, and a
drag proportional to the local density times the relative velocity, $n_\psi(v_\psi-v_d)$,
which vanishes when the condensate co-moves with the electron gas. The third term appears only
in the driven--dissipative case and feeds the pump--decay imbalance $\Gamma n_\psi$ into
the drag. 

We simulate the geometry shown in Fig.~\ref{fig:wire}, in which the non-resonant pump is spatially localized near one end of the wire. The exciton cloud generated at the pump spot produces a repulsive potential $U_R n_R$ and as a result the condensate feels a force, causing it to move ballistically along the wire with a well-defined momentum. To obtain the momentum distribution $N(k)$ of the polaritons, we Fourier transform the condensate wavefunction, using a spatial window that selects the flow region away from the pump spot; the parameters and further details of the numerics are given in the Appendix. The electron velocity enters the drag potential in Eq.~\eqref{eq:vdrag_relative_1d}. As shown in Fig.~\ref{fig:N_k}, the condensate momentum is shifted in response to the electron drift velocity. Thus, the microscopically derived potential gives the expected drag response: the condensate momentum is shifted in the direction of the electron flow. When the electron current co-propagates with the condensate flow, the electrons transfer momentum to the condensate shifting its momentum to a larger $k$; when the current counter-propagates, the condensate momentum is dragged shifting its momentum toward $k=0$, while the zero-current case lies between the two. Both the direction and magnitude of the momentum shift agree closely with the measurements of Ref.~\cite{Myers2025}, reproducing the experimentally observed co- and counter-propagating momentum distributions. The weaker peaks, present in both the measured and the simulated distributions, arise from reflection of the condensate off the boundary of the wire. This demonstrates that the microscopic reservoir interaction derived above naturally gives rise to the observed polariton drag without introducing an additional phenomenological drag term.

We find that an electron drift velocity of
$\pm 0.14~\mu\mathrm{m/ps}$ gives the best fit to the
momentum shift observed in the experimental data.
Ref.~\cite{Myers2025} estimated a velocity of
$10^6$ cm/s, i.e., $0.01~\mu\mathrm{m/ps}$,
based on fits to a quantum Boltzmann equation analysis.
This was considered a reasonable low-temperature estimate
compared with the room-temperature value of
$7.5\times10^{4}$ cm/s inferred from mobility measurements.

The present theory provides a more complete description
of the system than the model used in Ref.~\cite{Myers2025}.
In particular, the model of Ref.~\cite{Myers2025}
did not include pumping and decay and effectively treated
the polariton lifetime as infinite, whereas the present
model explicitly accounts for the open-dissipative nature
of the system by incorporating pumping and decay.
In addition, quantum--Boltzmann approaches evolve only
the occupation numbers and become inaccurate when
off-diagonal coherence terms become important, whereas
the present Gross--Pitaevskii-based treatment explicitly
describes the coherent condensate field.
Given these differences in the level of description,
a remaining difference of roughly one order of magnitude
between the electron velocity inferred in
Ref.~\cite{Myers2025} and the value used in the present
simulations is not unexpected.

The electron mobility in GaAs is known to increase
as the temperature is dropped from room temperature,
as phonon scattering is suppressed.
For example, in the modulation-doped GaAs samples
of Pfeiffer \textit{et al.}~\cite{pfeiffer1989electron},
the mobility increases from
$6.25\times10^{3}~\mathrm{cm^2/V\,s}$ at 295~K
to approximately $3\times10^{6}~\mathrm{cm^2/V\,s}$
at 10~K, i.e., by a factor of about $5\times10^{2}$.
Applying this factor to the room-temperature velocity
estimate gives approximately $0.375~\mu\mathrm{m/ps}$.
Importantly, the sample used in Ref.~\cite{Myers2025}
was also grown by Pfeiffer and coworkers, so this
provides a particularly relevant comparison for the
expected low-temperature mobility.

\section*{Conclusion}

We have derived a microscopic theory of drag for a Bose--Einstein condensate interacting with a moving reservoir. Tracing out a drifting fermionic bath within the Born--Markov approximation yields an effective G-P equation in which, within the coherent-field approximation, the influence of the reservoir is contained in a single real, history-dependent potential determined by the reservoir density correlator and the microscopic coupling $g$, without introducing an additional phenomenological drag coefficient. The resulting potential separates naturally into a static screening term that renormalizes the interaction and a drag term proportional to the relative velocity between the condensate and the reservoir. The reservoir-induced drag depends on the relative motion of the condensate and reservoir, opposes their velocity difference, and therefore drives the condensate toward co-motion with the reservoir. The drag potential is real-valued and therefore conserves the condensate number. Integrating this potential against the condensate density yields an effective drag force---a friction linear in the relative velocity---whose drag coefficient is fixed by the reservoir density fluctuations and the condensate form factor.

Applied to a one-dimensional exciton--polariton wire interacting with a dilute drifting electron gas, we insert the microscopically derived electron-drag potential into a standard phenomenological open-dissipative G-P model. The resulting simulations reproduce the observed drag: a current co-propagating with the condensate flow shifts the momentum distribution to higher $k$, while a counter-propagating current shifts it back toward lower $k$, in agreement with experiment. Since the microscopic derivation of the drag potential assumes only a coherent macroscopic field interacting with a reservoir, it connects directly to the energy-damping (scattering) reservoir interaction of the stochastic projected Gross--Pitaevskii theory of atomic condensates~\cite{gardiner2003stochastic,bradley2014spinor}, providing an experimentally accessible realization of the same number-conserving momentum-transfer mechanism. The same framework applies to atomic condensates coupled to thermal clouds, exciton condensates interacting with free carriers, and, more generally, quantum fluids interacting with moving reservoirs.

\section*{Acknowledgments}
The work carried out at the University of Pittsburgh was supported by the National Science Foundation under Grant No.~DMR-2306977. AB was supported by the Dodd-Walls Centre for Photonic and Quantum Technologies.

\bibliography{drag_refs}

@article{Narozhny2016,
  title   = {{Coulomb} drag},
  author  = {Narozhny, B. N. and Levchenko, A.},
  journal = {Rev. Mod. Phys.},
  volume  = {88},
  pages   = {025003},
  year    = {2016},
  doi     = {10.1103/RevModPhys.88.025003}
}

@article{Liu2017,
  title   = {Quantum {Hall} drag of exciton condensate in graphene},
  author  = {Liu, Xiaomeng and Watanabe, Kenji and Taniguchi, Takashi and Halperin, Bertrand I. and Kim, Philip},
  journal = {Nat. Phys.},
  volume  = {13},
  pages   = {746},
  year    = {2017},
  doi     = {10.1038/nphys4116}
}

@article{Nguyen2025,
  title   = {Perfect {Coulomb} drag in a dipolar excitonic insulator},
  author  = {Nguyen, Phuong X. and Ma, Liguo and Chaturvedi, Raghav and Watanabe, Kenji and Taniguchi, Takashi and Shan, Jie and Mak, Kin Fai},
  journal = {Science},
  volume  = {388},
  pages   = {274},
  year    = {2025},
  doi     = {10.1126/science.adl1829}
}

@article{Astrakharchik2004,
  title   = {Motion of a heavy impurity through a {Bose--Einstein} condensate},
  author  = {Astrakharchik, G. E. and Pitaevskii, L. P.},
  journal = {Phys. Rev. A},
  volume  = {70},
  pages   = {013608},
  year    = {2004},
  doi     = {10.1103/PhysRevA.70.013608}
}

@article{Sykes2009,
  title   = {Drag force on an impurity below the superfluid critical velocity in a quasi-one-dimensional {Bose--Einstein} condensate},
  author  = {Sykes, A. G. and Davis, M. J. and Roberts, D. C.},
  journal = {Phys. Rev. Lett.},
  volume  = {103},
  pages   = {085302},
  year    = {2009},
  doi     = {10.1103/PhysRevLett.103.085302}
}

@article{Berman2010,
  title   = {Drag effects in a system of electrons and microcavity polaritons},
  author  = {Berman, Oleg L. and Kezerashvili, Roman Ya. and Lozovik, Yurii E.},
  journal = {Phys. Rev. B},
  volume  = {82},
  pages   = {125307},
  year    = {2010},
  doi     = {10.1103/PhysRevB.82.125307}
}

@article{Ronning2020,
  title   = {Classical analogies for the force acting on an impurity in a {Bose--Einstein} condensate},
  author  = {R{\o}nning, Jonas and Skaugen, Audun and Hern{\'a}ndez-Garc{\'i}a, Emilio and L{\'o}pez, Crist{\'o}bal and Angheluta, Luiza},
  journal = {New J. Phys.},
  volume  = {22},
  pages   = {073018},
  year    = {2020},
  doi     = {10.1088/1367-2630/ab95de}
}

@article{carusotto2013quantum,
  title={Quantum fluids of light},
  author={Carusotto, Iacopo and Ciuti, Cristiano},
  journal={Reviews of Modern Physics},
  volume={85},
  number={1},
  pages={299--366},
  year={2013},
  publisher={APS}
}

@article{Wouters2007PRL,
  author  = {Michiel Wouters and Iacopo Carusotto},
  title   = {Excitations in a Nonequilibrium {Bose--Einstein} Condensate of Exciton Polaritons},
  journal = {Phys. Rev. Lett.},
  volume  = {99},
  pages   = {140402},
  year    = {2007},
  doi     = {10.1103/PhysRevLett.99.140402}
}

@article{Wouters2008SM,
  author  = {Michiel Wouters and Iacopo Carusotto},
  title   = {Excitations and superfluidity in non-equilibrium {Bose--Einstein} condensates of exciton-polaritons},
  journal = {Superlattices and Microstructures},
  volume  = {43},
  pages   = {524--527},
  year    = {2008},
  doi     = {10.1016/j.spmi.2007.07.024}
}

@article{Cotlet2019,
  title   = {Transport of Neutral Optical Excitations Using Electric Fields},
  author  = {Cotle\c{t}, Ovidiu and Pientka, Falko and Schmidt, Richard
             and Zar\'and, Gergely and Demler, Eugene and Imamo\u{g}lu, Ata\c{c}},
  journal = {Phys. Rev. X},
  volume  = {9},
  number  = {4},
  pages   = {041019},
  year    = {2019},
  doi     = {10.1103/PhysRevX.9.041019}
}

@article{tan2020interacting,
  title={Interacting polaron-polaritons},
  author={Tan, Li Bing and Cotlet, Ovidiu and Bergschneider, Andrea and Schmidt, Richard and Back, Patrick and Shimazaki, Yuya and Kroner, Martin and {\.I}mamo{\u{g}}lu, Ata{\c{c}}},
  journal={Physical Review X},
  volume={10},
  number={2},
  pages={021011},
  year={2020},
  publisher={APS}
}

@article{deng2002condensation,
  title={Condensation of {S}emiconductor {M}icrocavity {E}xciton {P}olaritons},
  author={Deng, Hui and Weihs, Gregor and Santori, Charles and Bloch, Jacqueline and Yamamoto, Yoshihisa},
  journal={Science},
  volume={298},
  number={5591},
  pages={199--202},
  year={2002},
  publisher={American Association for the Advancement of Science}
}

@article{kasprzak2006bose,
  title={{Bose--Einstein} {C}ondensation of {E}xciton {P}olaritons},
  author={Kasprzak, Jacek and Richard, Murielle and Kundermann, S and Baas, A and Jeambrun, P and Keeling, Jonathan Mark James and Marchetti, FM and Szyma{\'n}ska, MH and Andr{\'e}, R and Staehli, JL and others},
  journal={Nature},
  volume={443},
  number={7110},
  pages={409--414},
  year={2006},
  publisher={Nature Publishing Group}
}

@article{balili2007bose,
  title={{Bose--Einstein} {C}ondensation of {M}icrocavity {P}olaritons in a {T}rap},
  author={Balili, Ryan and Hartwell, V and Snoke, David and Pfeiffer, L and West, Kayte},
  journal={Science},
  volume={316},
  number={5827},
  pages={1007--1010},
  year={2007},
  publisher={American Association for the Advancement of Science}
}

@article{abbarchi2013macroscopic,
  title={Macroscopic {Q}uantum {S}elf-trapping and {Josephson} {O}scillations of {E}xciton {P}olaritons},
  author={Abbarchi, Marco and Amo, A and Sala, VG and Solnyshkov, DD and Flayac, H and Ferrier, L and Sagnes, I and Galopin, E and Lema{\^\i}tre, A and Malpuech, G and others},
  journal={Nature Physics},
  volume={9},
  number={5},
  pages={275--279},
  year={2013},
  publisher={Nature Publishing Group UK London}
}

@article{sanvitto2010persistent,
  title={Persistent {C}urrents and {Q}uantized {V}ortices in a {P}olariton {S}uperfluid},
  author={Sanvitto, D and Marchetti, FM and Szyma{\'n}ska, MH and Tosi, Guilherme and Baudisch, M and Laussy, Fabrice P and Krizhanovskii, DN and Skolnick, MS and Marrucci, L and Lemaitre, A and others},
  journal={Nature Physics},
  volume={6},
  number={7},
  pages={527--533},
  year={2010},
  publisher={Nature Publishing Group}
}

@article{lagoudakis2009observation,
  title={Observation of {H}alf-quantum {V}ortices in an {E}xciton-polariton {C}ondensate},
  author={Lagoudakis, KG and Ostatnick{\`y}, T and Kavokin, AV and Rubo, Yuri G and Andr{\'e}, R{\'e}gis and Deveaud-Pl{\'e}dran, Benoit},
  journal={Science},
  volume={326},
  number={5955},
  pages={974--976},
  year={2009},
  publisher={American Association for the Advancement of Science}
}

@article{Chervy2020,
  title   = {Accelerating Polaritons with External Electric and Magnetic Fields},
  author  = {Chervy, Thibault and Kn{\"u}ppel, Patrick and Abbaspour, Hadis
             and Lupatini, Mirko and F{\"a}lt, Stefan and Wegscheider, Werner
             and Kroner, Martin and Imamo{\u{g}}lu, Atac},
  journal = {Physical Review X},
  volume  = {10},
  number  = {1},
  pages   = {011040},
  year    = {2020},
  doi     = {10.1103/PhysRevX.10.011040}
}

@article{Meppelink2009,
  author  = {Meppelink, R. and Koller, S. B. and Vogels, J. M.
             and van der Straten, P. and Stoof, H. T. C.},
  title   = {Damping of Superfluid Flow by a Thermal Cloud},
  journal = {Physical Review Letters},
  volume  = {103},
  pages   = {265301},
  year    = {2009},
  doi     = {10.1103/PhysRevLett.103.265301}
}

@article{Nandi2012,
  author  = {Nandi, D. and Finck, A. D. K. and Eisenstein, J. P.
             and Pfeiffer, L. N. and West, K. W.},
  title   = {Exciton Condensation and Perfect {Coulomb} Drag},
  journal = {Nature},
  volume  = {488},
  number  = {7412},
  pages   = {481--484},
  year    = {2012},
  doi     = {10.1038/nature11302}
}

@article{Myers2025,
  author  = {Myers, D. M. and Yao, Q. and Alnatah, H. and Mukherjee, S.
             and Ozden, B. and Beaumariage, J. and Pfeiffer, L. N.
             and West, K. and Snoke, D. W.},
  title   = {Pushing Photons with Electrons: Observation of the
             Polariton Drag Effect},
  journal = {Physical Review Letters},
  volume  = {135},
  number  = {14},
  pages   = {146903},
  year    = {2025},
  doi     = {10.1103/fb2r-qdq7}
}

@article{li2021theory,
  title={Theory of polariton-electron interactions in semiconductor microcavities},
  author={Li, Guangyao and Bleu, Olivier and Levinsen, Jesper and Parish, Meera M},
  journal={Physical Review B},
  volume={103},
  number={19},
  pages={195307},
  year={2021},
  publisher={APS}
}

@article{ramon2002scattering,
  title={Scattering of polaritons by a two-dimensional electron gas in a semiconductor microcavity},
  author={Ramon, G and Rapaport, R and Qarry, A and Cohen, E and Mann, A and Ron, Arza and Pfeiffer, LN},
  journal={Physical Review B},
  volume={65},
  number={8},
  pages={085323},
  year={2002},
  publisher={APS}
}

@article{wouters2010superfluidity,
  title={Superfluidity and critical velocities in nonequilibrium {Bose--Einstein} condensates},
  author={Wouters, Michiel and Carusotto, Iacopo},
  journal={Physical Review Letters},
  volume={105},
  number={2},
  pages={020602},
  year={2010},
  publisher={APS}
}

@article{alnatah2024coherence,
  title={Coherence measurements of polaritons in thermal equilibrium reveal a power law for two-dimensional condensates},
  author={Alnatah, Hassan and Yao, Qi and Beaumariage, Jonathan and Mukherjee, Shouvik and Tam, Man Chun and Wasilewski, Zbigniew and West, Ken and Baldwin, Kirk and Pfeiffer, Loren N and Snoke, David W},
  journal={Science Advances},
  volume={10},
  number={18},
  pages={eadk6960},
  year={2024},
  publisher={American Association for the Advancement of Science}
}

@article{alnatah2024bose,
  title={{Bose--Einstein} condensation of polaritons at room temperature in a {GaAs/AlGaAs} structure},
  author={Alnatah, Hassan and Liang, Shuang and Yao, Qi and Wan, Qiaochu and Beaumariage, Jonathan and West, Ken and Baldwin, Kirk and Pfeiffer, Loren N and Snoke, David W},
  journal={{ACS} Photonics},
  volume={12},
  number={1},
  pages={48--52},
  year={2024},
  publisher={ACS Publications}
}

@article{sun2017bose,
  title={{Bose--Einstein} condensation of long-lifetime polaritons in thermal equilibrium},
  author={Sun, Yongbao and Wen, Patrick and Yoon, Yoseob and Liu, Gangqiang and Steger, Mark and Pfeiffer, Loren N and West, Ken and Snoke, David W and Nelson, Keith A},
  journal={Physical Review Letters},
  volume={118},
  number={1},
  pages={016602},
  year={2017},
  publisher={APS}
}

@article{steger2015slow,
  title={Slow {R}eflection and {T}wo-photon {G}eneration of {M}icrocavity {E}xciton--polaritons},
  author={Steger, Mark and Gautham, Chitra and Snoke, David W and Pfeiffer, Loren and West, Ken},
  journal={Optica},
  volume={2},
  number={1},
  pages={1--5},
  year={2015},
  publisher={Optica Publishing Group}
}

@inproceedings{ravets2019nonlinear,
  title={Nonlinear optics in the fractional quantum {Hall} regime},
  author={Ravets, Sylvain},
  booktitle={International Conference on New Trends in Quantum Light and Nanophysics ({QLIN}) 2019},
  year={2019}
}

@article{sidler2017fermi,
  title={{Fermi} polaron-polaritons in charge-tunable atomically thin semiconductors},
  author={Sidler, Meinrad and Back, Patrick and Cotlet, Ovidiu and Srivastava, Ajit and Fink, Thomas and Kroner, Martin and Demler, Eugene and Imamoglu, Atac},
  journal={Nature Physics},
  volume={13},
  number={3},
  pages={255--261},
  year={2017},
  publisher={Nature Publishing Group UK London}
}

@article{bobrovska2015adiabatic,
  title={Adiabatic approximation and fluctuations in exciton-polariton condensates},
  author={Bobrovska, Nataliya and Matuszewski, Micha{\l}},
  journal={Physical Review B},
  volume={92},
  number={3},
  pages={035311},
  year={2015},
  publisher={APS}
}

@article{alnatah2024critical,
  title={Critical fluctuations in a confined driven-dissipative quantum condensate},
  author={Alnatah, Hassan and Comaron, Paolo and Mukherjee, Shouvik and Beaumariage, Jonathan and Pfeiffer, Loren N and West, Ken and Baldwin, Kirk and Szyma{\'n}ska, Marzena and Snoke, David W},
  journal={Science Advances},
  volume={10},
  number={12},
  pages={eadi6762},
  year={2024},
  publisher={American Association for the Advancement of Science}
}

@article{fontaine2022kardar,
  title={{Kardar--Parisi--Zhang} universality in a one-dimensional polariton condensate},
  author={Fontaine, Quentin and Squizzato, Davide and Baboux, Florent and Amelio, Ivan and Lema{\^\i}tre, Aristide and Morassi, Martina and Sagnes, Isabelle and Le Gratiet, Luc and Harouri, Abdelmounaim and Wouters, Michiel and others},
  journal={Nature},
  volume={608},
  number={7924},
  pages={687--691},
  year={2022},
  publisher={Nature Publishing Group UK London}
}

@article{comaron2018dynamical,
  title={Dynamical critical exponents in driven-dissipative quantum systems},
  author={Comaron, P and Dagvadorj, G and Zamora, A and Carusotto, I and Proukakis, NP and Szyma{\'n}ska, MH},
  journal={Physical Review Letters},
  volume={121},
  number={9},
  pages={095302},
  year={2018},
  publisher={APS}
}

@book{breuer2002theory,
  title={The theory of open quantum systems},
  author={Breuer, Heinz-Peter and Petruccione, Francesco},
  year={2002},
  publisher={OUP Oxford}
}

@book{carmichael2013statistical,
  title={Statistical methods in quantum optics 1: master equations and {Fokker--Planck} equations},
  author={Carmichael, Howard J},
  year={2013},
  publisher={Springer Science \& Business Media}
}

@article{kubo1966fluctuation,
  title={The fluctuation-dissipation theorem},
  author={Kubo, Rep},
  journal={Reports on Progress in Physics},
  volume={29},
  number={1},
  pages={255--284},
  year={1966}
}

@article{callen1951irreversibility,
  title={Irreversibility and generalized noise},
  author={Callen, Herbert B and Welton, Theodore A},
  journal={Physical Review},
  volume={83},
  number={1},
  pages={34},
  year={1951},
  publisher={APS}
}

@article{frisch1992transition,
  title={Transition to dissipation in a model of superflow},
  author={Frisch, Thomas and Pomeau, Yves and Rica, Sergio},
  journal={Physical Review Letters},
  volume={69},
  number={11},
  pages={1644},
  year={1992},
  publisher={APS}
}

@article{pavloff2002breakdown,
  title={Breakdown of superfluidity of an atom laser past an obstacle},
  author={Pavloff, Nicolas},
  journal={Physical Review A},
  volume={66},
  number={1},
  pages={013610},
  year={2002},
  publisher={APS}
}

@article{amo2009superfluidity,
  title={Superfluidity of polaritons in semiconductor microcavities},
  author={Amo, Alberto and Lefr{\`e}re, J{\'e}r{\^o}me and Pigeon, Simon and Adrados, Claire and Ciuti, Cristiano and Carusotto, Iacopo and Houdr{\'e}, Romuald and Giacobino, Elisabeth and Bramati, Alberto},
  journal={Nature Physics},
  volume={5},
  number={11},
  pages={805--810},
  year={2009},
  publisher={Nature Publishing Group UK London}
}

@article{yao2025persistent,
  title={Persistent, controllable circulation of a polariton ring condensate},
  author={Yao, Qi and Comaron, Paolo and Alnatah, Hassan and Beaumariage, Jonathan and Mukherjee, Shouvik and West, Ken and Pfeiffer, Loren and Baldwin, Kirk and Szyma{\'n}ska, Marzena H and Snoke, David},
  journal={Optica},
  volume={12},
  number={7},
  pages={991--996},
  year={2025},
  publisher={Optica Publishing Group}
}

@article{gardiner2003stochastic,
  title   = {The stochastic {G}ross--{P}itaevskii equation: {II}},
  author  = {Gardiner, C. W. and Davis, M. J.},
  journal = {J. Phys. B: At. Mol. Opt. Phys.},
  volume  = {36},
  number  = {23},
  pages   = {4731},
  year    = {2003},
  doi     = {10.1088/0953-4075/36/23/010}
}

@article{bradley2014spinor,
  title   = {Stochastic projected {G}ross-{P}itaevskii equation for spinor and multicomponent condensates},
  author  = {Bradley, A. S. and Blakie, P. B.},
  journal = {Phys. Rev. A},
  volume  = {90},
  number  = {2},
  pages   = {023631},
  year    = {2014},
  doi     = {10.1103/PhysRevA.90.023631}
}

@article{Rooney2012SPGPE,
  title = {Stochastic projected Gross-Pitaevskii equation},
  author = {Rooney, S. J. and Blakie, P. B. and Bradley, A. S.},
  journal = {Phys. Rev. A},
  volume = {86},
  issue = {5},
  pages = {053634},
  year = {2012},
  doi = {10.1103/PhysRevA.86.053634}
}

@article{Mehdi2023MutualFriction,
  title = {Mutual friction and diffusion of two-dimensional quantum vortices},
  author = {Mehdi, Zain and Hope, Joseph J. and Szigeti, Stuart S. and Bradley, Ashton S.},
  journal = {Phys. Rev. Res.},
  volume = {5},
  issue = {1},
  pages = {013184},
  year = {2023},
  doi = {10.1103/PhysRevResearch.5.013184}
}

@article{Krause2024JRSolitons,
  title = {Thermal decay of planar Jones-Roberts solitons},
  author = {Krause, Nils A. and Bradley, Ashton S.},
  journal = {Phys. Rev. A},
  volume = {110},
  issue = {5},
  pages = {053302},
  year = {2024},
  doi = {10.1103/PhysRevA.110.053302}
}

@article{blakie2008dynamics,
  title={Dynamics and statistical mechanics of ultra-cold Bose gases using c-field techniques},
  author={Blakie, P Blair and Bradley, AS and Davis, MJ and Ballagh, RJ and Gardiner, CW},
  journal={Advances in Physics},
  volume={57},
  number={5},
  pages={363--455},
  year={2008},
  publisher={Taylor \& Francis}
}

@book{Annett2004,
  author    = {James F. Annett},
  title     = {Superconductivity, Superfluids, and Condensates},
  series    = {Oxford Master Series in Condensed Matter Physics},
  number    = {5},
  publisher = {Oxford University Press},
  address   = {Oxford},
  year      = {2004},
  isbn      = {978-0-19-850756-7}
}

@book{NozieresPines1990,
  author    = {Philippe Nozi{\`e}res and David Pines},
  title     = {Theory of Quantum Liquids, Volume {II}: Superfluid {B}ose Liquids},
  publisher = {Addison-Wesley},
  address   = {Redwood City, CA},
  year      = {1990},
  isbn      = {978-0-201-40841-6}
}

@book{LifshitzPitaevskii1980,
  author    = {E. M. Lifshitz and L. P. Pitaevskii},
  title     = {Statistical Physics, Part 2: Theory of the Condensed State},
  series    = {Course of Theoretical Physics},
  volume    = {9},
  publisher = {Pergamon Press},
  address   = {Oxford},
  year      = {1980},
  isbn      = {978-0-08-023073-7}
}

@article{alnatah2025strong,
  title={Strong coupling of polaritons at room temperature in a GaAs/AlGaAs structure},
  author={Alnatah, Hassan and Liang, Shuang and Wan, Qiaochu and Beaumariage, Jonathan and West, Ken and Baldwin, Kirk and Pfeiffer, Loren N and Tam, Man Chun Alan and Wasilewski, Zbigniew R and Snoke, David W},
  journal={Physical Review B},
  volume={112},
  number={4},
  pages={045307},
  year={2025},
  publisher={APS}
}

@article{pfeiffer1989electron,
  title={Electron Mobilities Exceeding $10^7$ cm$^2$/V s in Modulation-Doped GaAs},
  author={Pfeiffer, Loren and West, K. W. and Stormer, H. L. and Baldwin, K. W.},
  journal={Applied Physics Letters},
  volume={55},
  number={18},
  pages={1888--1890},
  year={1989},
  publisher={American Institute of Physics}
}

\onecolumngrid
\newpage
\medskip

\appendix

\section{Derivation of the master equation}
\label{app:master}

Equation~\eqref{eq:ME} follows from the second-order term of Eq.~\eqref{eq:iter}.
Tracing over the reservoir and applying the Born approximation,
$\tilde\rho(t')\simeq\tilde\rho_\psi(t')\otimes\rho_\phi$, the boson dynamics at order
$g^2$ are
\begin{equation}
\frac{\partial\tilde\rho_\psi(t)}{\partial t}
=-\frac{1}{\hbar^2}\int_0^t\! dt'\,
\Tr_\phi\big[\tilde H_{\rm int}(t),
[\tilde H_{\rm int}(t'),\tilde\rho_\psi(t')\otimes\rho_\phi]\big],
\label{eq:app_start}
\end{equation}
with $\tilde H_{\rm int}(t)=g\int d^Dr\,\tilde n_\psi(\rr,t)\,\delta\tilde n_\phi(\rr,t)$,
the reservoir correlator
$C(\rr-\rr',\tau)=\langle\delta\tilde n_\phi(\rr,t)\,\delta\tilde n_\phi(\rr',t')\rangle$
with ($\tau=t-t'$) and, by the Hermiticity of $\hat n_\phi$,
$\langle\delta\tilde n_\phi(\rr',t')\,\delta\tilde n_\phi(\rr,t)\rangle=C^*(\rr-\rr',\tau)$.

Expanding the commutator,
\begin{align*}
&\big[\tilde H_{\rm int}(t),[\tilde H_{\rm int}(t'),\tilde\rho_\psi(t')\otimes\rho_\phi]\big]
\notag\\
&\quad=
\tilde H_{\rm int}(t)\,\tilde H_{\rm int}(t')\,\tilde\rho_\psi(t')\otimes\rho_\phi
-\tilde H_{\rm int}(t)\,\big(\tilde\rho_\psi(t')\otimes\rho_\phi\big)\,\tilde H_{\rm int}(t')
\notag\\
&\quad\phantom{=}
-\tilde H_{\rm int}(t')\,\big(\tilde\rho_\psi(t')\otimes\rho_\phi\big)\,\tilde H_{\rm int}(t)
+\big(\tilde\rho_\psi(t')\otimes\rho_\phi\big)\,\tilde H_{\rm int}(t')\,\tilde H_{\rm int}(t).
\end{align*}
Because $\tilde n_\psi$ and $\delta\tilde n_\phi$ act on different Hilbert spaces and
$\tilde\rho_\psi(t')\otimes\rho_\phi$ is a product state, each term factorizes; tracing
over the reservoir with cyclicity of $\Tr_\phi$ turns every reservoir factor into $C$ or
$C^*$,
\begin{align*}
&\Tr_\phi\big[\tilde H_{\rm int}(t),[\tilde H_{\rm int}(t'),\tilde\rho_\psi(t')\otimes\rho_\phi]\big]
=g^2\!\int\! d^Dr\,d^Dr'\,\Big\{
\notag\\
&\quad
C(\rr-\rr',\tau)\,\tilde n_\psi(\rr,t)\,\tilde n_\psi(\rr',t')\,\tilde\rho_\psi(t')
-C^*(\rr-\rr',\tau)\,\tilde n_\psi(\rr,t)\,\tilde\rho_\psi(t')\,\tilde n_\psi(\rr',t')
\notag\\
&\quad
-C(\rr-\rr',\tau)\,\tilde n_\psi(\rr',t')\,\tilde\rho_\psi(t')\,\tilde n_\psi(\rr,t)
+C^*(\rr-\rr',\tau)\,\tilde\rho_\psi(t')\,\tilde n_\psi(\rr',t')\,\tilde n_\psi(\rr,t)
\Big\}.
\end{align*}
The four terms combine pairwise into commutators,
\begin{align}
&\Tr_\phi\big[\tilde H_{\rm int}(t),[\tilde H_{\rm int}(t'),\tilde\rho_\psi(t')\otimes\rho_\phi]\big]
=g^2\!\int\! d^Dr\,d^Dr'\,\Big\{
\notag\\
&\quad
C(\rr-\rr',\tau)\,\big[\tilde n_\psi(\rr,t),\,\tilde n_\psi(\rr',t')\,\tilde\rho_\psi(t')\big]
-C^*(\rr-\rr',\tau)\,\big[\tilde n_\psi(\rr,t),\,\tilde\rho_\psi(t')\,\tilde n_\psi(\rr',t')\big]
\Big\}.
\label{eq:app_group}
\end{align}
Since $\tilde n_\psi$ and $\tilde\rho_\psi$ are Hermitian,
$\big[\tilde n_\psi(\rr,t),\tilde n_\psi(\rr',t')\tilde\rho_\psi(t')\big]^\dagger
=-\big[\tilde n_\psi(\rr,t),\tilde\rho_\psi(t')\tilde n_\psi(\rr',t')\big]$, so the second
term in Eq.~\eqref{eq:app_group} is the Hermitian conjugate of the first:
\begin{align}
&\Tr_\phi\big[\tilde H_{\rm int}(t),[\tilde H_{\rm int}(t'),\tilde\rho_\psi(t')\otimes\rho_\phi]\big]
\notag\\
&\quad
=g^2\!\int\! d^Dr\,d^Dr'\,
\Big\{C(\rr-\rr',\tau)\,\big[\tilde n_\psi(\rr,t),\,\tilde n_\psi(\rr',t')\,\tilde\rho_\psi(t')\big]
+\mathrm{h.c.}\Big\}.
\label{eq:app_hc}
\end{align}
Inserting Eq.~\eqref{eq:app_hc} into Eq.~\eqref{eq:app_start}, applying the Markov approximation $\tilde\rho_\psi(t')\to\tilde\rho_\psi(t)$, and changing variables to $\tau=t-t'$, we obtain Eq.~\eqref{eq:ME} after extending the upper integration limit to infinity. This extension is justified because the reservoir correlation functions decay rapidly on the timescale $\tau_c$, so the integrand is negligible for $\tau\gtrsim\tau_c$.

\section{Transformation of the master equation to the Schr\"odinger picture}
\label{app:schroedinger}

We start from the interaction-picture master equation \eqref{eq:ME} and go back to the Schr\"odinger-picture given by Eq.~\eqref{eq:MES}. We use the fact that unitary conjugation distributes over products, and therefore over commutators: for any unitary $U$,
\begin{align}
U^\dagger (AB)\, U&=(U^\dagger A U)(U^\dagger B U)
\notag\\
\Longrightarrow\qquad
U^\dagger [A,B]\, U&=[\,U^\dagger A U,\;U^\dagger B U\,],
\label{eq:homo}
\end{align}
which follows by inserting $1=UU^\dagger$ between every pair of operators. 

The interaction picture was defined with respect to $H_0=H_\psi+H_\phi$. Since
$[H_\psi,H_\phi]=0$, the evolution operator factorizes,
$e^{iH_0t/\hbar}=e^{iH_\psi t/\hbar}e^{iH_\phi t/\hbar}$, and since $H_\phi$ commutes
with every boson operator,
\begin{align}
\tilde n_\psi(\rr,t)
&=e^{iH_0t/\hbar}\,\hat n_\psi(\rr)\,e^{-iH_0t/\hbar}
\notag\\
&=e^{iH_\psi t/\hbar}\,\hat n_\psi(\rr)\,e^{-iH_\psi t/\hbar}.
\label{eq:nfree}
\end{align}
After the reservoir has been traced out, the boson density matrices in the two
pictures are therefore related by
\begin{equation}
\rho_\psi(t)=e^{-iH_\psi t/\hbar}\,\tilde\rho_\psi(t)\,e^{iH_\psi t/\hbar}.
\label{eq:pict}
\end{equation}

Differentiating \eqref{eq:pict}, we obtain
\begin{align}
\frac{\partial\rho_\psi}{\partial t}
={}&-\frac i\hbar H_\psi\, e^{-iH_\psi t/\hbar}\tilde\rho_\psi e^{iH_\psi t/\hbar}
\notag\\
&+e^{-iH_\psi t/\hbar}\tilde\rho_\psi e^{iH_\psi t/\hbar}\,\frac i\hbar H_\psi
\notag\\
&+e^{-iH_\psi t/\hbar}\,\frac{\partial\tilde\rho_\psi}{\partial t}\,e^{iH_\psi t/\hbar}
\notag\\[2pt]
={}&-\frac i\hbar\big[H_\psi,\rho_\psi\big]
+e^{-iH_\psi t/\hbar}\,\frac{\partial\tilde\rho_\psi}{\partial t}\,e^{iH_\psi t/\hbar}.
\label{eq:step1}
\end{align}

We next insert the right-hand side of \eqref{eq:ME} into the last term of \eqref{eq:step1}. The
correlation function $C$ is a c-number and passes through the conjugation; by
\eqref{eq:homo} the conjugation acts on each operator of the nested product separately.
With $U=e^{iH_\psi t/\hbar}$ and using \eqref{eq:nfree} and \eqref{eq:pict}:
\begin{align}
&e^{-iH_\psi t/\hbar}\,\tilde n_\psi(\rr,t)\,e^{iH_\psi t/\hbar}
\notag\\
&\quad=e^{-iH_\psi t/\hbar}\Big(e^{iH_\psi t/\hbar}\hat n_\psi(\rr)e^{-iH_\psi t/\hbar}\Big)
e^{iH_\psi t/\hbar}
\notag\\
&\quad=\hat n_\psi(\rr),
\label{eq:conj1}\\[4pt]
&e^{-iH_\psi t/\hbar}\,\tilde n_\psi(\rr',t-\tau)\,e^{iH_\psi t/\hbar}
\notag\\
&\quad=e^{-iH_\psi t/\hbar}\Big(e^{iH_\psi (t-\tau)/\hbar}\hat n_\psi(\rr')
e^{-iH_\psi(t-\tau)/\hbar}\Big)e^{iH_\psi t/\hbar}
\notag\\
&\quad=e^{-iH_\psi\tau/\hbar}\,\hat n_\psi(\rr')\,e^{iH_\psi\tau/\hbar}
\;\equiv\;\hat n_\psi(\rr',-\tau),
\label{eq:conj2}\\[4pt]
&e^{-iH_\psi t/\hbar}\,\tilde\rho_\psi(t)\,e^{iH_\psi t/\hbar}
=\rho_\psi(t).
\label{eq:conj3}
\end{align}
Finally, using \eqref{eq:conj1}--\eqref{eq:conj3} inside the commutator of \eqref{eq:ME}
(allowed by \eqref{eq:homo}),
\begin{align*}
&e^{-iH_\psi t/\hbar}\,
\big[\tilde n_\psi(\rr,t),\,\tilde n_\psi(\rr',t-\tau)\,\tilde\rho_\psi(t)\big]\,
e^{iH_\psi t/\hbar}
\notag\\
&\qquad=
\big[\hat n_\psi(\rr),\,\hat n_\psi(\rr',-\tau)\,\rho_\psi(t)\big],
\end{align*}
and identically for the hermitian conjugate. Substituting into \eqref{eq:step1}
reproduces Eq.~\eqref{eq:MES}.

\section{Alternative form of the drag potential}
\label{app:altform}

In this section, we integrate by parts to rewrite Equation~\eqref{eq:vdrag_moments} in a form that makes the physical origin of the drag more transparent. Starting from
\begin{align}
V_{\rm drag}(\mathbf x,t)
={}&
\frac{2g^2}{\hbar}
\int d^{D}\rr''\!\int\!\frac{d^{D}\q}{(2\pi)^{D}}
\Big\{
\operatorname{Im}\!\big[e^{i\q\cdot(\mathbf x-\mathbf r'')}K_0(\q)\big]\,
n_\psi(\mathbf r'',t)
\notag\\
&-
\operatorname{Im}\!\big[e^{i\q\cdot(\mathbf x-\mathbf r'')}K_1(\q)\big]\,
(\partial_t+\mathbf v_d\!\cdot\!\nabla'')\,n_\psi(\mathbf r'',t)
\Big\},
\label{eq:Vdrag_appendix_start}
\end{align}
and substituting the continuity equation $\partial_t n_\psi=-\nabla\cdot\mathbf j_\psi$, we obtain
\begin{equation*}
-(\partial_t+\mathbf v_d\cdot\nabla)n_\psi
=\nabla\cdot\mathbf j_\psi-\mathbf v_d\cdot\nabla n_\psi .
\end{equation*}
Since the reservoir drift velocity is spatially uniform,
$\mathbf v_d\cdot\nabla n_\psi=\nabla\cdot(\mathbf v_d n_\psi)$, we can write
\begin{equation}
-(\partial_t+\mathbf v_d\cdot\nabla)n_\psi
=\nabla\cdot(\mathbf j_\psi-\mathbf v_d n_\psi).
\label{eq:relative_current}
\end{equation}
Substituting Eq.~\eqref{eq:relative_current} into
Eq.~\eqref{eq:Vdrag_appendix_start} gives
\begin{align*}
V_{\rm drag}(\mathbf x,t)
={}&
\frac{2g^2}{\hbar}
\int d^{D}\rr''\!\int\!\frac{d^{D}\q}{(2\pi)^{D}}
\Big\{
\operatorname{Im}\!\big[e^{i\q\cdot(\mathbf x-\mathbf r'')}K_0(\q)\big]\,
n_\psi
\notag\\
&+
\operatorname{Im}\!\big[e^{i\q\cdot(\mathbf x-\mathbf r'')}K_1(\q)\big]\,
\nabla''\!\cdot\!(\mathbf j_\psi-\mathbf v_d n_\psi)
\Big\}.
\end{align*}
Integrating the second term by parts,
\begin{align*}
\int d^{D}\rr''\,A\,\nabla''\!\cdot\!\mathbf J
={}&
\int_{\partial\Omega}d\mathbf S''\cdot A\mathbf J
\notag\\
&-\int d^{D}\rr''\,(\nabla''A)\cdot\mathbf J,
\end{align*}
with $A=\operatorname{Im}[e^{i\q\cdot(\mathbf x-\mathbf r'')}K_1]$ and
$\mathbf J=\mathbf j_\psi-\mathbf v_d n_\psi$. Assuming periodic boundaries, or 
the wavefunction of the condensate vanishes at hard walls so that the surface term vanishes,
and using
\begin{align*}
\nabla''\operatorname{Im}\!\big[e^{i\q\cdot(\mathbf x-\mathbf r'')}K_1\big]
&=\operatorname{Im}\!\big[-i\q\,e^{i\q\cdot(\mathbf x-\mathbf r'')}K_1\big]
\notag\\
&=-\q\,\operatorname{Re}\!\big[e^{i\q\cdot(\mathbf x-\mathbf r'')}K_1\big],
\end{align*}
the drag potential becomes
\begin{align}
V_{\rm drag}(\mathbf x,t)
={}&
\frac{2g^2}{\hbar}
\int d^{D}\rr''\!\int\!\frac{d^{D}\q}{(2\pi)^{D}}
\Big\{
\operatorname{Im}\!\big[e^{i\q\cdot(\mathbf x-\mathbf r'')}K_0(\q)\big]\,
n_\psi(\mathbf r'',t)
\notag\\
&+
\q\cdot\big[\mathbf j_\psi(\mathbf r'',t)-\mathbf v_d n_\psi(\mathbf r'',t)\big]\,
\operatorname{Re}\!\big[e^{i\q\cdot(\mathbf x-\mathbf r'')}K_1(\q)\big]
\Big\}.
\label{eq:Vdrag_relative_current}
\end{align}
Finally, using $\mathbf j_\psi=n_\psi\mathbf v_\psi$, the current difference becomes
$\mathbf j_\psi-\mathbf v_d n_\psi=n_\psi(\mathbf v_\psi-\mathbf v_d)$, so that
Eq.~\eqref{eq:Vdrag_relative_current} depends only on the relative motion between the
condensate and the reservoir.

\section{Relative-current form of the one-dimensional drag kernel}
\label{app:relcurrent}

The one-dimensional drag potential for the exciton-polariton case, Eq.~\eqref{eq:vdrag_1d}, can be rewritten in a form that explicitly separates the number-conserving terms from the gain and loss terms. We define
\begin{equation*}
\Gamma(x,t)
=
\frac{RP(x,t)}{\gamma_R+Rn_\psi(x,t)}
-\gamma_c ,
\end{equation*}
and
\begin{equation*}
\mathcal F(x-x'')
=
\ln\!\left(\frac{L}{\pi\ell_T}\right)
-\frac{\gamma_E}{2}
-2\!\int_0^{|x-x''|/\ell_T}\!\!\!D(s)\,ds .
\end{equation*}
Using Eq.~\eqref{eq:comoving_driven},
\begin{equation*}
(\partial_t+v_d\partial_{x''})n_\psi
=
-\partial_{x''}j_\psi
+
v_d\partial_{x''}n_\psi
+
\Gamma n_\psi .
\end{equation*}
Since the drift velocity is spatially uniform, we can write
\begin{equation*}
-\partial_{x''}j_\psi
+
v_d\partial_{x''}n_\psi
=
\partial_{x''}(v_dn_\psi-j_\psi),
\end{equation*}
so that the second term of Eq.~\eqref{eq:vdrag_1d} becomes
\begin{equation*}
\frac{g^{2}n_e\ell_T}{\sqrt{\pi}\,\hbar v_T^{2}}
\int dx''\,
\mathcal F(x-x'')
\left[
\partial_{x''}(v_dn_\psi-j_\psi)
+
\Gamma n_\psi
\right].
\end{equation*}
Integrating the transport term by parts gives
\begin{align*}
&\int dx''\,
\mathcal F(x-x'')\,
\partial_{x''}(v_dn_\psi-j_\psi)
\notag\\
={}&
\big[
\mathcal F(x-x'')\,(v_dn_\psi-j_\psi)
\big]_{\rm boundaries}
\notag\\
&-
\int dx''\,
\partial_{x''}\mathcal F(x-x'')\,(v_dn_\psi-j_\psi).
\end{align*}
The derivative of the kernel is
\begin{equation*}
\partial_{x''}\mathcal F(x-x'')
=
\frac{2}{\ell_T}\,
D\!\left(\frac{x-x''}{\ell_T}\right),
\end{equation*}
Like before, we assume hard-wall boundaries so that the boundary terms vanish and the drag potential becomes
\begin{align*}
V_{\rm drag}(x,t)
={}&
-\frac{g^{2}n_e}{\hbar v_T}\sqrt{\frac{\pi}{2}}
\int dx''\,
\operatorname{erfc}\!\left(\frac{|x-x''|}{\ell_T}\right)
n_\psi(x'',t)
\notag\\
&+\frac{2g^{2}n_e}{\sqrt{\pi}\,\hbar v_T^{2}}
\int dx''\,
D\!\left(\frac{x-x''}{\ell_T}\right)
\big[\,j_\psi(x'',t)-v_dn_\psi(x'',t)\,\big]
\notag\\
&+\frac{g^{2}n_e\ell_T}{\sqrt{\pi}\,\hbar v_T^{2}}
\int dx''\,
\mathcal F(x-x'')\,
\Gamma(x'',t)\,n_\psi(x'',t),
\end{align*}
which is the relative-current form used in Eq.~\eqref{eq:vdrag_relative_1d}.

\section{Numerical parameters and integration scheme}
\label{app:numerics}

Table~\ref{tab:params} gives the parameters used in the simulations shown in Fig.~\ref{fig:N_k}.  The polariton-polariton and polariton-electron interactions are written in terms of the bare exciton--exciton interaction $U_0$, the exciton fraction $|X|^2$, the number of quantum wells $N^{\mathrm{QW}}$, and the
wire width $W$; the exciton fraction enters as $|X|^4$ for the polariton--polariton interaction and as $|X|^2$ for the polariton--electron interaction. The interaction strengths were divided by the number of quantum wells since the separation of the quantum wells is much larger than the Bohr radius of the exciton, implying that excitons in adjacent quantum wells do not interact. The reservoir enters the drag only through the single combination
$g^2 n_e$ of the polariton--electron coupling and the one-dimensional electron density, which
we therefore list directly.

\begin{table}[h]
\begin{ruledtabular}
\begin{tabular}{lll}
Quantity & Symbol & Value \\
\hline
Polariton mass            & $m_\psi$      & $1\times10^{-4}\,m_0$ \\
Electron mass             & $m_\phi$      & $0.067\,m_0$ \\
Exciton fraction          & $|X|^2$       & $0.5$ \\
Bare exciton interaction  & $U_0$         & $40\ \mu$eV\,$\mu$m$^2$ \\
Number of quantum wells   & $N^{\mathrm{QW}}$ & $12$ \\
Wire width                & $W$           & $10\ \mu$m \\
Polariton--polariton int.  & $U$   & $\dfrac{|X|^4 U_0}{W N^{\mathrm{QW}}}$ \\[6pt]
Polariton--reservoir int.  & $U_R$ & $\dfrac{|X|^2 U_0}{W N^{\mathrm{QW}}}$ \\[6pt]
Drag coupling             & $g^2 n_e$     & $42\ \mu$eV$^2\,\mu$m \\
Electron temperature      & $T$           & $10$ K \\
Polariton lifetime        & $\gamma_c^{-1}$ & $270$ ps \\
Reservoir decay rate      & $\gamma_R$    & $0.1$ ps$^{-1}$ \\
Stimulated scattering     & $R$           & $5\times10^{-3}\ \mu$m\,ps$^{-1}$ \\
Pump amplitude            & $P_0$         & $1.5\ \mu$m$^{-1}$ps$^{-1}$ \\
Pump position             & $x_p$         & $-77\ \mu$m \\
Pump FWHM                 & $w_p$         & $7.7\ \mu$m \\
Electron drift            & $v_d$         & $0,\ \pm0.14\ \mu$m/ps \\
\end{tabular}
\end{ruledtabular}
\caption{Parameters of the one-dimensional polariton-wire simulations.}
\label{tab:params}
\end{table}

Hard wall boundaries were imposed at the two ends of the wire. The pump term $P(x)$ in Eq.~\eqref{eq:reservoir_dynamics_numerical} is a Gaussian centered at $x_p$ near one end of the wire, written directly in terms of its full width at half maximum $w_p$,
\begin{equation*}
P(x)=P_0\,\exp\!\left[-4\ln 2\,\frac{(x-x_p)^2}{w_p^{2}}\right].
\end{equation*}

The open-dissipative G-P equation~\eqref{eq:effective_driven_GPE} is integrated on a uniform grid of $N=510$ points with spacing of $\Delta x=0.6\ \mu$m. Spatial derivatives are evaluated by centered finite differences. Time stepping uses a fourth-order Runge--Kutta scheme; the runs extend to $\approx 5$ ns, which is long compared with the polariton lifetime and is sufficient to reach a steady state.

Momentum distributions are obtained from the Fourier transform of $\psi$ over the forward-flow region $x\in[-62,31]\ \mu$m, taken with a rectangular window away from the pump spot. $N(k)=|\psi(k)|^{2}$ is averaged over the steady state and normalized to unit peak height. The velocity of the electrons was used as a fit parameter to obtain the best fit to the momentum shifts in the experimental data.

\end{document}